\documentclass[twocolumn,tighten]{aastex62}
\pdfoutput=1
\usepackage{array}
\usepackage{amsmath}
\usepackage{multirow}
\usepackage{graphicx}
\usepackage{placeins}
\usepackage[utf8]{inputenc}
\usepackage[percent]{overpic}
\usepackage{color}

\begin{document}

\def\fesc{\textit{f$_{esc}$}}
\def\hst{\textit{HST}}
\def\uvmag{F275W$_{\mathrm{AB}}$}
\def\zmag{\textit{z}$_{\mathrm{850}}$}

\title{$z \sim 2.5 - 3$ Ionizers in the GOODS-N Field}

\author[0000-0002-1706-7370]{L.~H.~Jones}
\affiliation{Department of Astronomy, University of Wisconsin-Madison,
475 N. Charter Street, Madison, WI 53706, USA}

\author[0000-0002-3306-1606]{A.~J.~Barger}
\affiliation{Department of Astronomy, University of Wisconsin-Madison,
475 N. Charter Street, Madison, WI 53706, USA}
\affiliation{Department of Physics and Astronomy, University of Hawaii,
2505 Correa Road, Honolulu, HI 96822, USA}
\affiliation{Institute for Astronomy, University of Hawaii, 2680 Woodlawn Drive,
Honolulu, HI 96822, USA}

\author[0000-0002-6319-1575]{L.~L.~Cowie}
\affiliation{Institute for Astronomy, University of Hawaii,
2680 Woodlawn Drive, Honolulu, HI 96822, USA}

\author[0000-0001-5851-6649]{P.~Oesch}
\affiliation{Geneva Observatory, Universit{\'e} de Gen{\`e}ve, 
Chemin des Maillettes 51, 1290 Versoix, Switzerland}

\author{E.~M.~Hu}
\affiliation{Institute for Astronomy, University of Hawaii,
2680 Woodlawn Drive, Honolulu, HI 96822, USA}

\author{A.~Songaila}
\affiliation{Institute for Astronomy, University of Hawaii,
2680 Woodlawn Drive, Honolulu, HI 96822, USA}

\author[0000-0003-3997-5705]{R. P. Naidu}
\affiliation{Harvard-Smithsonian Center for Astrophysics, 
60 Garden Street, Cambridge, MA 02138, USA}

\begin{abstract}
We use deep F275W imaging from the \textit{Hubble} Deep UV Legacy Survey (HDUV) and G280 grism spectroscopy from \textit{HST}/WFC3, along with new and archival optical spectra from Keck/DEIMOS, to search for candidate ionizing sources in the GOODS-N field at $z \sim$ 2.5 -- 3. Spectroscopic identification of our UV-selected sources are 99\% complete to F275W = 25.5 in the region of the UV imaging, and we identify 6 potential ionizing galaxies or AGNs at $z \sim$ 3.  By far the brightest of these is a $z = 2.583$ AGN that totally dominates the ionizing flux in the region, with a specific ionizing volume emissivity at 912 \AA~of $\epsilon_{912}$ = $8.3^{27}_{1.4} \times 10^{24}$~erg~s$^{-1}$~Hz$^{-1}$~Mpc$^{-3}$. Based on our spectroscopic data, we find four candidates are contaminated by foreground galaxies at $z \sim$ 0.5 -- 0.7. At $\epsilon_{912}$ = $2.2^{7.2}_{0.4} \times 10^{23}$~erg~s$^{-1}$~Hz$^{-1}$~Mpc$^{-3}$, the remaining candidate galaxy's contribution to the ionizing background lies well below the flux required to ionize the intergalactic medium at $z \sim$ 2.5 -- 3, consistent with previous observations that show AGNs provide the bulk of the ionizing background at these redshifts. 
\end{abstract}

\keywords{cosmology: observations 
--- galaxies: active --- galaxies: distances and redshifts 
--- galaxies: evolution --- galaxies: formation}

\section{Introduction}
\label{intro}
One of the most pressing issues in modern observational cosmology is the identification of sources that contribute to the metagalactic ionizing background, particularly in the era of cosmic reionization --- an important epoch in the history of the Universe that saw the formation of the first stars and galaxies at $z \gtrsim$ 6 (e.g., \citealt{bou06,bou12,bou15,ouch09,rob15}). Star-forming galaxies and active galactic nuclei (AGNs) both contribute to the production of ionizing photons, though their relative importance appears to evolve with cosmic time. Most evidence currently favors a scenario in which low-luminosity star-forming galaxies are the primary driver of hydrogen reionization \citep{ric00,bou06,font07,font14,rob10,rob15,jap17}, while AGN contributions to the ionizing background are small until $z \sim$ 2 -- 3 \citep{barg03,bolt05,cbt09,cri16,smith17,puch18}. However, some authors have argued that quasars/AGNs could remain important at very high redshifts, producing a non-negligible or even dominant fraction of UV photons during the era of reionization (e.g., \citealt{font12,gia15}, \citealt{mad15}).

One motivation for these latter studies is to relax constraints on the escape fraction, $f_{esc}$, needed to produce the observed ionizing background at high redshift; these constraints are imposed by a faint galaxy dominated reionization scenario. Indeed, determining a value for $f_{esc}$, which is the fraction of all Lyman continuum (LyC, rest frame $\lambda$ $<$ 912 \AA{}) photons that manage to escape their galaxy of origin to ionize the intergalactic medium (IGM), has been a major focal point of research on reionization. Most theoretical and semi-analytical models of reionization require an average $f_{esc}$ of about 10\% or greater for star-forming galaxies (e.g., \citealt{bolt07,van12a,feng16,price16,kimm17}; see, however, \citealt{fauch08} and \citealt{matt17}), though at the highest redshifts $f_{esc}$ remains largely unconstrained by observations. For sources at $z \gtrsim$ 4, the low transmissivity of the IGM effectively prohibits direct measurements of $f_{esc}$ \citep{mad95,song04,in14}. Thus, observations focused on analogous objects at slightly lower redshifts are used to constrain the ionization history of the Universe.

Previous individual detections or stacked data analyses suggest small values of $f_{esc}$ in the local universe, at most $\sim1-3\%$ (e.g., \citealt{lei95,stei01,grim09,cow10,leit13,rut16}), with some indications that the escape fraction increases with decreasing UV luminosity and/or increasing redshift (e.g., \citealt{mit13,font14,fai16,kai16,jap17}).
Significant object-to-object variance and differences in the average $f_{esc}$ between types of sources (i.e., $f_{esc} \gtrsim$ 0.5 for AGNs versus a few percent for galaxies) further complicates the quest for a reliable measurement of the global ionizing escape fraction \citep{fern03,ma15,cri16,graz16,gua16}.

Much effort has thus been expended in building up a statistically significant population of LyC-emitting sources across a range of redshifts. A handful of strong LyC emitters have been detected in the local universe using data from the Sloan Digital Sky Survey (SDSS), the COS spectrograph on the \textit{Hubble Space Telescope} (\hst{}), and other facilities (e.g., \citealt{berg13,bort14,izo16a,izo16b,izo18,lei16}). Additional individual detections at redshifts $\sim$ 2 -- 3 have been made, albeit with some contamination from foreground objects (e.g., \citealt{van10a,van10b,van12b,most15,siana15,graz16,shap16}), while stacking analyses tend to give a relatively weak average LyC signal at $z \gtrsim 3$ (e.g., \citealt{marchi17,rut17,nai18,stei18}).

The GOODS-North and South fields \citep{goods04} are particularly attractive 
targets for LyC-emitter searches due to the abundance of ancillary data, including thorough spectroscopic coverage. For example, \citet{cbt09} (hereafter CBT09) used a sample of X-ray selected broad-line AGNs in the GOODS-N to estimate the contributions of AGNs and galaxies to the ionizing background over $0 < z < 5$. They found a $2\sigma$ upper limit of 0.008 for the escape fraction for galaxies at $z \sim 1.15$ and, interestingly, that the AGN contribution at similar redshifts is dominated by a small number of far-UV (FUV)-bright quasars. \citet{siana10} used \textit{HST} imaging of the GOODS fields to search for LyC emission at $z \sim$ 1.3 and constrain the relative escape fraction ($f_{esc,rel}$, the LyC flux relative to the UV continuum flux, typically at 1500 \AA). They found a stacked upper limit of $f_{esc,rel} <$ 0.02 with no galaxies in their sample detected individually. More recently, \citet{graz17} used \textit{U}- and \textit{R}-band imaging from the Large Binocular Telescope to constrain the escape of LyC photons at $z \sim$ 3.3 in several deep fields, including the GOODS-N field, and found $f_{esc,rel}$ is at most 1.7\% for their stacked image of 69 star-forming galaxies.
Meanwhile, a particularly strong LyC candidate at $z \sim$ 3.2 in the GOODS-S, known as \textit{Ion2}, was discovered by \citet{van15} and later confirmed by \citet{van16} and \citet{deb16}, who found it to be a compact, low-metallicity source with an absolute escape fraction upwards of 50\%. 

\citet{nai17} identified another six candidate LyC sources in the GOODS fields at $z \sim$ 2 (all with $f_{esc} \gtrsim$ 13\%) using \hst{}/WFC3 imaging in the F275W and F336W bands from the \textit{Hubble} Deep UV (HDUV) Legacy Survey (GO13872; \citealt{oesch18}). At the redshifts probed by \citet{nai17}, the Lyman break lies at $\sim$2750 \AA{}, such that both ionizing and non-ionizing photons fall within the F275W window. To determine the true contribution of LyC photons to the F275W flux then requires somewhat sophisticated and correspondingly uncertain modeling, along with Monte Carlo simulations of UV color and IGM attenuation. At redshifts greater than $\sim$2.4, however, the F275W filter exclusively probes LyC photons, making the HDUV data a valuable asset for identifying ionizing sources at high redshifts.

In this paper, we combine new and preexisting optical spectroscopy on the GOODS-N field with the deep, high spatial resolution F275W data from the HDUV survey to obtain limits on the contributions of candidate LyC-emitting galaxies at $z\sim3$, where the HDUV filter set probes only the Lyman continuum, to the overall ionizing emissivity from star-forming galaxies and low-luminosity AGNs. We also present a new UV grism spectroscopic observation from \hst{}/WFC3 of a $z \sim 2.6$ FUV-bright quasar.

In Section \ref{data}, we describe the data we used to select and characterize possible high-redshift LyC emitters, including UV and X-ray imaging, optical spectra from Keck/DEIMOS, and G280 grism spectroscopy. In Section \ref{results}, we describe our search for candidate LyC emitters and discuss the properties of the sources we found, along with potential sources of contamination by foreground galaxies. In Section \ref{flux}, we estimate the associated contributions (or limits thereof) to the ionizing background at $z \sim$ 3 and compare to the flux required to maintain an ionized IGM at this redshift. In Section \ref{summary}, we summarize our findings and discuss future prospects for the field.

We assume $\Omega_{M}$ = 0.3, $\Omega_{\Lambda}$ = 0.7, and $H_{0}$ = 70~km~s$^{-1}$~Mpc$^{-1}$ throughout this work. All magnitudes are given in the AB system, defined as $m_{AB} = -2.5 \mathrm{log} f_{\nu} - 48.60$ for flux density, $f_{\nu}$, in units of erg~s$^{-1}$~cm$^{-2}$~Hz$^{-1}$.

\section{Data}
\label{data}

\subsection{F275W Imaging}
\label{datahduv}
The HDUV survey (GO13872; \citealt{oesch18}) is a 132-orbit WFC3 imaging 
program centered on the GOODS-North and South fields. Designed to capitalize on 
existing WFC3/UVIS imaging from the CANDELS \citep{grog11,koek11} and 
UVUDF \citep{tep13,rafel15} surveys, the HDUV survey imaged both of these fields 
in the F275W and F336W filters around or within the existing CANDELS and UVUDF 
footprints. When combined with imaging from each of these surveys, the reduced 
HDUV images achieve depths of $\approx 27.5$ and $27.9$ mag in the F275W 
and F336W filters, respectively ($5\sigma$ detection, 0\farcs4 diameter aperture). 
Since the Lyman continuum is redshifted into the F275W bandpass at $z > 2.4$, 
he deep and relatively wide F275W coverage provided by the HDUV survey enables 
us to search for potential sources of ionizing radiation at high redshift.

\subsection{Optical/NIR Spectroscopy}
\label{zspec}
Secure spectroscopic redshifts are required for reliable identification of candidate LyC emitters within our F275W sample. The GOODS-N field is one of the most heavily-studied regions of the sky, with a wealth of existing spectroscopic data from DEIMOS on Keck~I and LRIS and MOSFIRE on Keck~II (e.g., \citealt{coh00,cow04,cow16,swin04,wirth04,wirth15,chap05,red06,barg08,trou08,coop11,kriek15,u15,cow16}). We crossmatched our sample (defined in Section \ref{dataf275}) to existing Keck spectroscopic catalogs to determine redshifts, then used DEIMOS
to target any F275W source in our sample without existing spectroscopic identifications, 
or to obtain additional spectra of candidate LyC emitters to check for 
possible contamination by foreground galaxies (see Sections~\ref{gnuvc1} -- \ref{gnuvc6}).

For our new DEIMOS observations,
we used the 600~line~mm$^{-1}$ grating, giving a $d\lambda$ of 3.5~\AA\ and a wavelength coverage
of 5300~\AA. We centered the spectra at an average wavelength of 7200~\AA, but the exact
wavelength range for each spectrum depends on the position of the slit in the mask. We broke
each $\sim1$~hr exposure into three sub-exposures positioned at a central position and two 
offset positions stepped $1\farcs5$ in each direction along the slit.
Our dithering procedure provides extremely high-precision sky subtraction.
We reduced the spectra following the procedures described in \citet{cow96}.

\subsection{UV Grism Spectroscopy}
\label{grizspec}
The \hst{}/WFC3 grism spectrum from program
GO12479 (PI: Hu)
was based on 5 dithered observations with the G280 grism. 
Each observation was 475~s, giving a total exposure time of 2375~s.
We also obtained a 120~s imaging exposure with the F2000LP filter
to set the zero point for computing the shape of the spectrum and
the wavelength calibration. The G280 grism extends to a short
wavelength of 1900~\AA\ with a resolution of 70 at 3000~\AA,
giving coverage down to a rest wavelength of 530~\AA. 
We measured the flux from the first order spectrum using the
calibrations of the spatial distortion and wavelength relative
to the zeroth order given in \citet{kunt09}.
We extracted the spectrum as a function of wavelength with a 
6~pixel ($0\farcs24$) boxcar centered on the central position of the spectrum.
Finally, we flux calibrated the spectrum in units of microJansky,
though the absolute calibration is not critical in the present analysis.

\subsection{X-ray Imaging}
\label{xrays}
To identify probable AGNs in our F275W sample, we used X-ray data from the 2~Ms \textit{Chandra X-ray Observatory} exposure of the \textit{Chandra} Deep Field-North \citep{alex03,xue16}. This image reaches a limiting flux of $f_{0.5-2 \mathrm{ keV}} \approx 1.5 \times 10^{-17}$~erg~cm$^{-2}$~s$^{-1}$ near the central aim point. We used a 1\farcs5 search radius to identify X-ray counterparts to sources in our F275W sample; 60 had X-ray counterparts.
We computed the rest-frame 2 -- 8 keV luminosities, $L_{X}$, of these counterparts from the 0.5 -- 2 keV fluxes with an assumed $\Gamma = 1.8$ and no absorption correction using
\begin{equation}
L_X = 4\pi d^2_L f_{0.5-2 \mathrm{keV}}\bigg(\frac{1+z}{4}\bigg)^{\Gamma-2}\ \mathrm{erg\ s^{-1}.}
\end{equation}
We classify any source with an X-ray luminosity $L_{X} > 10^{44}$ erg s$^{-1}$ as a quasar (red squares enclosed by a purple open square in Figure~\ref{bzspec_fig}).

\section{Search for $\lowercase{z} \sim 3$ Candidate LyC Emitters}
\label{results}

\subsection{F275W Sample}
\label{dataf275}
We started with all \zmag{} $<$ 26 galaxies from the 
140~arcmin$^2$ GOODS-N observations of \citet{goods04} obtained 
with \hst{}'s Advanced Camera for Surveys (ACS). At $z \sim$ 3, the ACS 
F850LP filter probes the rest-frame FUV at $\sim$2300 \AA, providing a 
good selection of likely star-forming galaxies at these redshifts.

\begin{figure}[t]
\includegraphics[angle=0,width=3.35in]{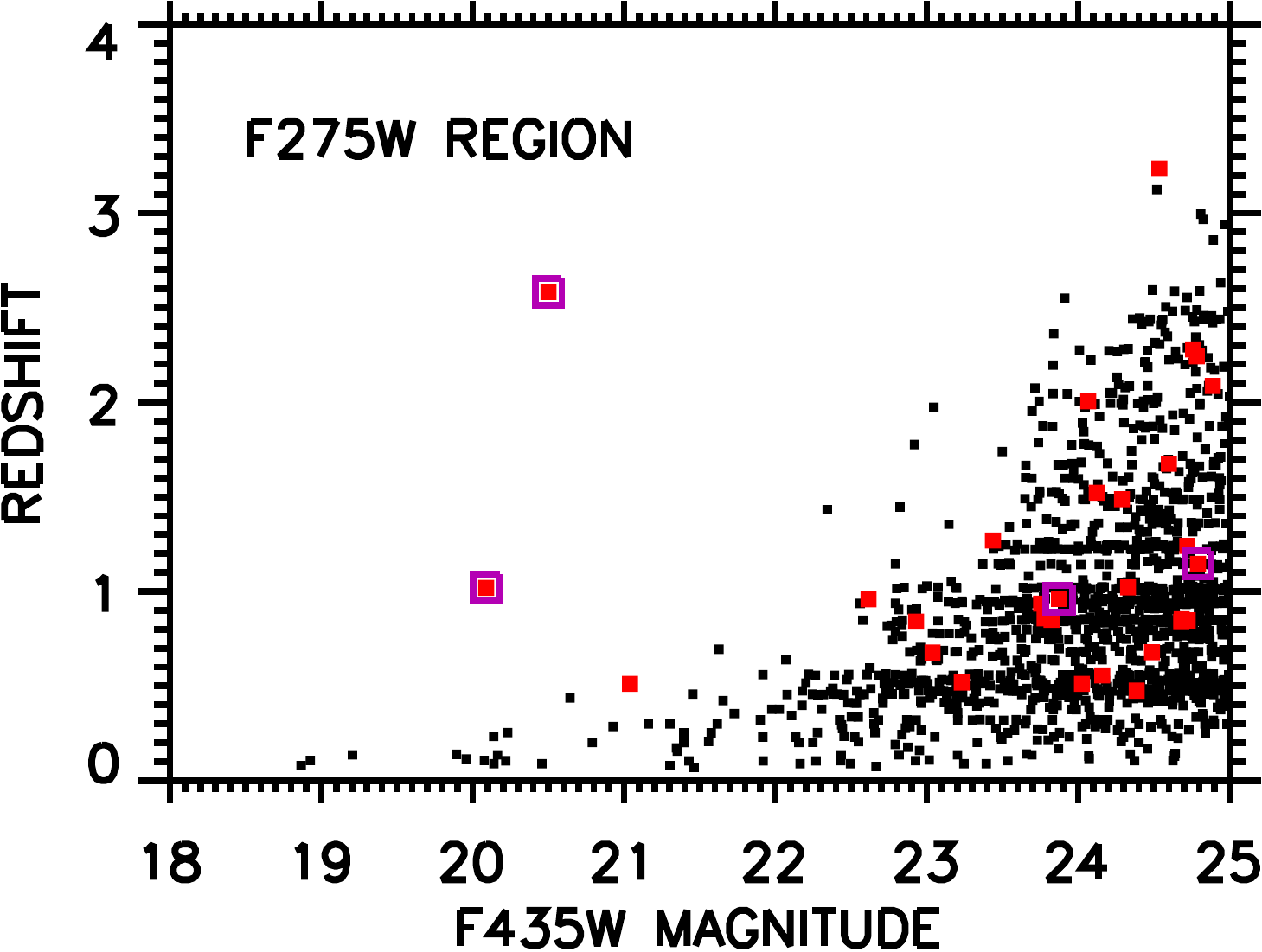}
\vspace{0.15cm}
\caption{Spectroscopic redshift vs. F435W $(B)$ magnitude
for the 68~arcmin$^2$ area covered by the HDUV GOODS-N F275W image. 
Sources with no X-ray counterpart are denoted by black squares, while sources 
with an X-ray detection are denoted by red squares, and those with quasar X-ray 
luminosities are enclosed in purple open squares. The spectroscopic 
identifications only start to become significantly incomplete (82\% identified) at 
B magnitudes of $24.5-25$.
\label{bzspec_fig}
}
\end{figure}

We then restricted to the 68~arcmin$^2$ area
where there is F275W coverage with rms errors fainter than 27~mag.
There are 5712 sources with \zmag{} $<$ 26 in this area.
In Figure~\ref{bzspec_fig}, we plot redshift versus F435W $(B)$ magnitude 
for this area. The spectroscopic identifications are essentially complete below 
$B = 24$ but drop to 95\% at $24-24.5$ and 82\% at $24.5-25$.

We next measured the F275W magnitudes within $1\arcsec$ diameter apertures 
at the positions of each \zmag{} $<$ 26 source using a customized
IDL routine and subtracting the background using the median
in a $3-6\arcsec$ annulus. 
Magnitude errors were measured from the associated rms noise files.
{\em We hereafter consider the 1063 sources with F275W magnitudes 
brighter than 26 (4$\sigma$) as our UV sample.}

\begin{figure}
\includegraphics[width=3.35in]{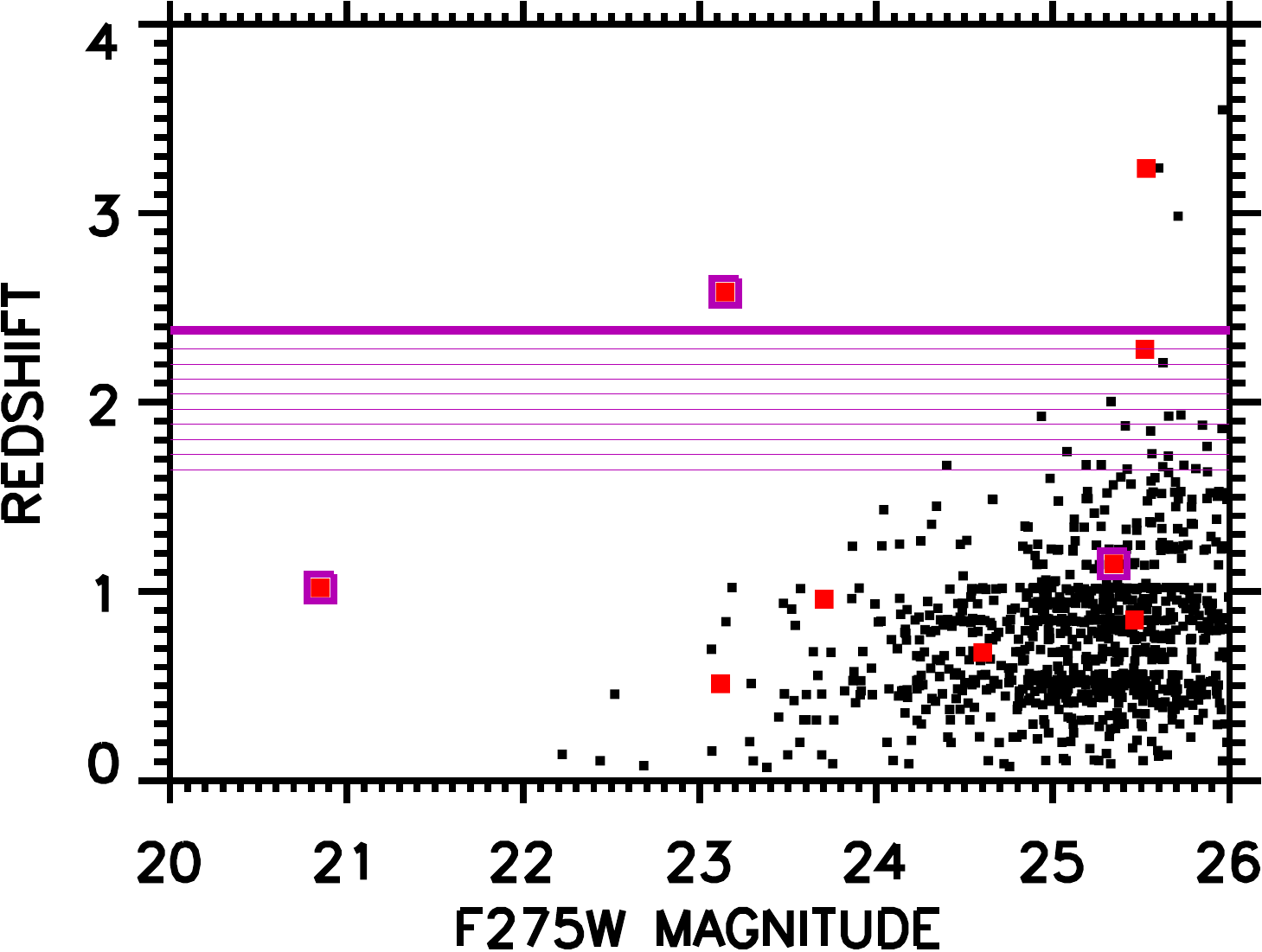}
\caption{Spectroscopic redshift vs. F275W magnitude 
for the 68~arcmin$^2$ area covered by the HDUV GOODS-N F275W image. 
Sources with no X-ray counterpart are denoted by black squares, while sources 
with an X-ray detection are denoted by red squares, and those with quasar X-ray 
luminosities are enclosed in purple open squares. 
The thick purple line marks the 
redshift above which the F275W filter is 
sampling solely below the Lyman continuum 
break ($z$ = 2.36). The purple hatched region shows 
the redshift range where the 
break falls within the filter bandpass.
The spectroscopic
identifications only start to become 
significantly incomplete (68$\%$ identified) 
at F275W magnitudes of $25.5-26$.
\label{zspec_fig}
}
\end{figure}

In Figure~\ref{zspec_fig}, we show redshift versus F275W magnitude for this sample.
The spectroscopic identifications are essentially 
complete below F275W = 25 but drop 
to 97\% at  F275W$ =  25-25.5$ and 68\% at $25.5-26$.
Of the 138 F275W $< 26$ sources without spectroscopic 
identifications, 38 have been observed but not identified.
The remaining sources  have not been observed.

Only five sources in Figure~\ref{zspec_fig} lie above the $z \sim$ 2.4 
threshold (thick purple line) where the F275W flux consists solely of LyC 
photons (assuming no contamination from foreground sources).
One of these sources is an X-ray AGN, and another is an X-ray quasar.

{\setlength{\extrarowheight}{3pt}
\begin{table*}[ht]
    \normalsize
    \centering
    \caption{\\\textsc{Summary of Six Candidate LyC Emitters}}
    \begin{tabular}{c c c c c c c c c c}
    \hline
    \hline
    ID$^{a}$ & R.A. & Dec. & $z_{spec}^{b}$ & $z_{grism}^{c}$ & $L_{X}$ & F275W & F435W$_{\mathrm{AB}}$ & F606W$_{\mathrm{AB}}$ & $f_{ion}^{d}$ \\
    \hline
    GN-UVC-1$^{1,2}$ & 189.095581 & +62.257492 & 2.583$^{I}$ & 2.597 & $3.44 \times 10^{44}$ & 23.14 & 20.50 & 20.49 & 0.087 \\
    
    GN-UVC-2$^{1}$ & 189.179535 & +62.185806 & 3.236$^{I}$ & 3.299  & $5.66 \times 10^{42}$ & 25.53 & 24.54 & 23.40 & --- \\
    
    GN-UVC-3$^{1,2}$ & 189.275543 & +62.250462 & 3.239$^{I,II}$ & --- & --- & 25.60 & 25.21 & 24.46 & --- \\
 
    GN-UVC-4$^{1}$ & 189.148758 & +62.271030 & 2.984$^{I,II}$ & --- & --- & 25.71 & 25.05 & 24.58 & --- \\
    
    GN-UVC-5$^{1}$ & 189.296936 & +62.270989 & 3.546$^{I,III}$ & --- & --- & 25.96 & 25.77 & 25.44 & --- \\
    
    GN-UVC-6$^{2}$ & 189.201889 & +62.266682 & 2.439$^{II}$ & --- & --- & 26.53 & 24.93 & 24.74 & 0.193 \\
    \hline
    \end{tabular}
\label{sources}
\tablecomments{
$^a$Superscripts indicate if a candidate was selected by (1) its F275W magnitude, (2) its F275W-F435W color, or both. \\
$^b$Spectroscopic redshifts from (I) this work; (II) \citet{red06}; and (III) \citet{u15}. \\
$^c$Determined from G280 grism data from \hst{}/WFC3 (GO12479, PI: Hu) for GN-UVC-1, and from G141 grism data from the 3D-HST survey \citep{3dhst} for GN-UVC-2. \\
$^d$Ratio of F275W flux to F606W flux (rest-frame $\sim675$ \AA~to 1500 \AA; see Section \ref{flux}).
}
\end{table*}
}

\subsection{Color-Selected Sample}
\label{colorselect}

\begin{figure}[t]
\includegraphics[trim=10 13 55 29,clip,width=3.35in]{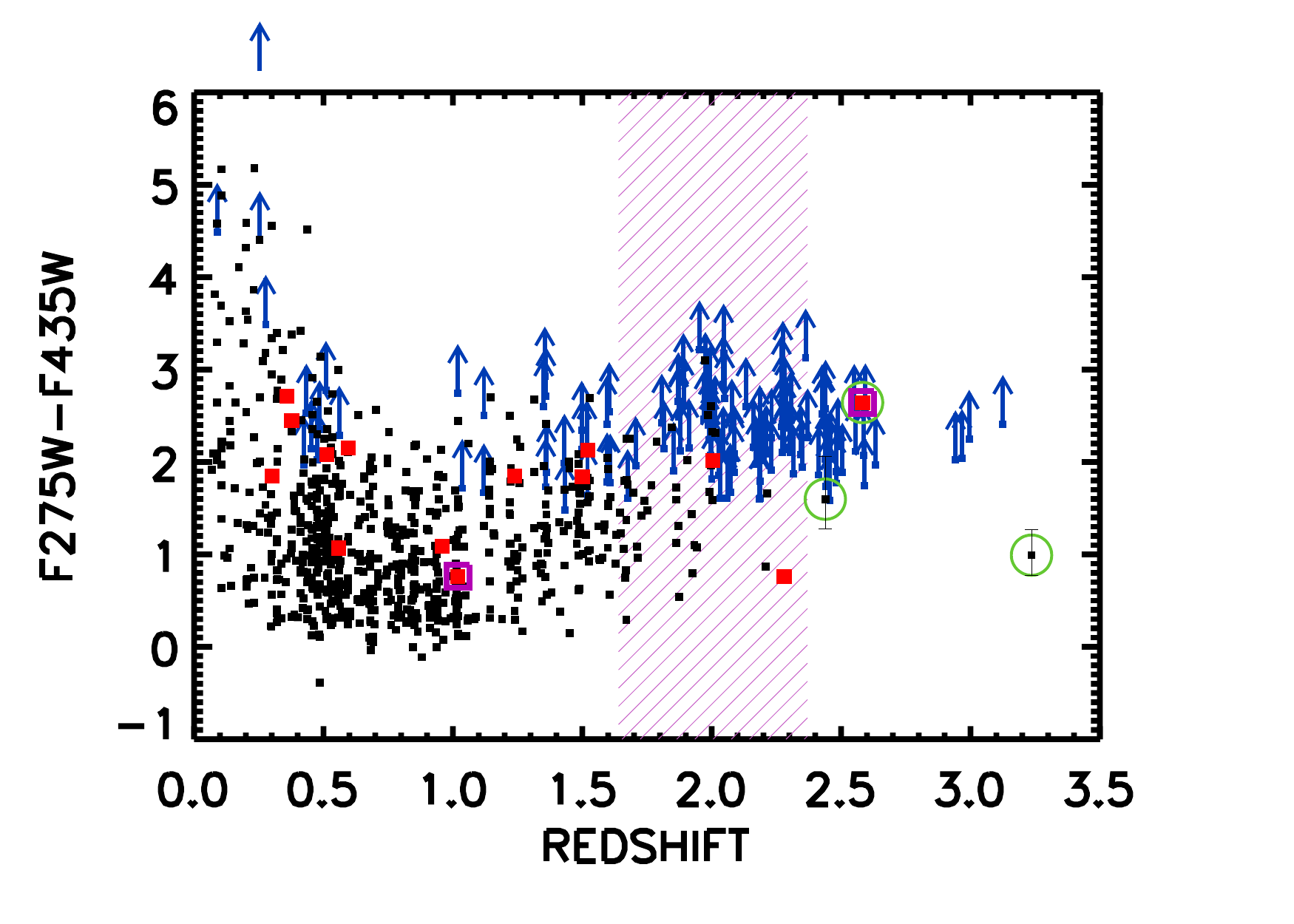}
\caption{Observed F275W$-$F435W vs. redshift for $B < 25$ galaxies with flat rest-frame UV continua (i.e., selected using $V$ -- \zmag{} $< 1$). Sources with no X-ray counterpart are denoted by black squares, while sources with an X-ray detection are denoted by red squares, and those with quasar X-ray luminosities are enclosed in purple open squares. Sources with lower limits on F275W$-$F435W are plotted at their $2\sigma$ values with blue upward pointing arrows. The purple hatched region marks the redshift range where the F275W filter straddles the Lyman break. The three sources enclosed in green open circles have measured F275W$-$F435W colors at $>2\sigma$ significance and are shown with error bars that reflect the 1$\sigma$ uncertainties in the F275W magnitudes.
\label{color_z}
}
\end{figure}

Alternatively, we can utilize a color selection to search for candidate ionizing sources. 
In order to have a substantially complete spectroscopic sample, we start with sources 
with $B < 25$ (see Figure~\ref{bzspec_fig}). 
We then use $V$ -- \zmag{} $< 1$ to select galaxies with relatively flat UV continua (i.e., likely star-forming galaxies) at high redshifts. We plot F275W$-$F435W versus redshift in Figure~\ref{color_z} for the sources that meet these criteria. We indicate with purple hatching
the redshift range where the F275W filter straddles the Lyman break. The typical color becomes noticeably redder around $z \sim 2$ as the LyC break moves into this window, with most objects at $z \gtrsim 2$ having so little F275W flux that we can measure only lower limits on the color. We find three $z > 2.36$ sources that have measured F275W-F435W colors at the $>2\sigma$ level. We show these with error bars and enclosed in green circles in Figure~\ref{color_z}. Two of these color-selected sources also fall into our F275W-selected sample (see Section~\ref{dataf275}) and appear in Figure~\ref{zspec_fig} (one is the X-ray quasar), 
while the third is detected at the 2.9$\sigma$ level in F275W.

\subsection{Six Candidate LyC Emitters}
\label{lyccandidates}
In Table \ref{sources}, we list the basic properties of our six candidate LyC emitters, including ID number, decimal coordinates, ground-based spectroscopic and (when available) {\em HST\/} grism redshifts, X-ray luminosities, F275W, $B$, and $V$ magnitudes, and ionization fraction $f_{ion}$ (see Section \ref{flux}).
We show in Figure \ref{pascal_thumbs} both the F275W thumbnail (left) and three-color thumbnail (right; red = F160W, green = F606W, blue = F435W) images of each source. 
In the following subsections, we briefly discuss for each
of the six sources individually our efforts to try and confirm the LyC emission from
the $z\sim3$ sources.

\begin{figure*}[t]
\setcounter{figure}{3}
\centering
\begin{tabular}{p{2.62in}p{2.62in}}
  
  \begin{overpic}[trim=140 230 92 197,clip,width=2.6in,height=2.6in]{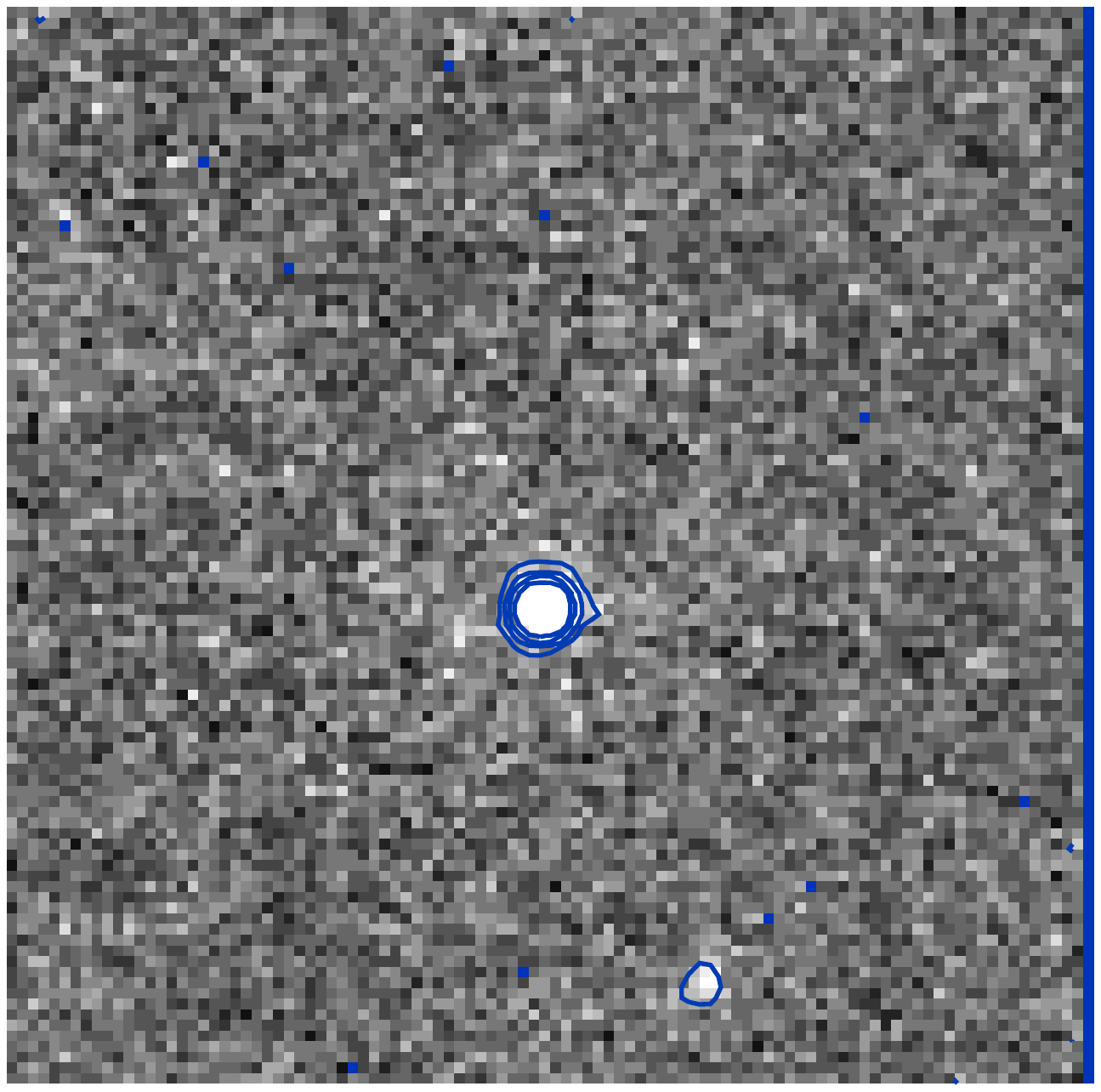} \put(7,87){\large \color{white} \textbf{GN-UVC-1}}\end{overpic} &
  \includegraphics[trim=118 247 118 190,clip,width=2.6in,height=2.6in]{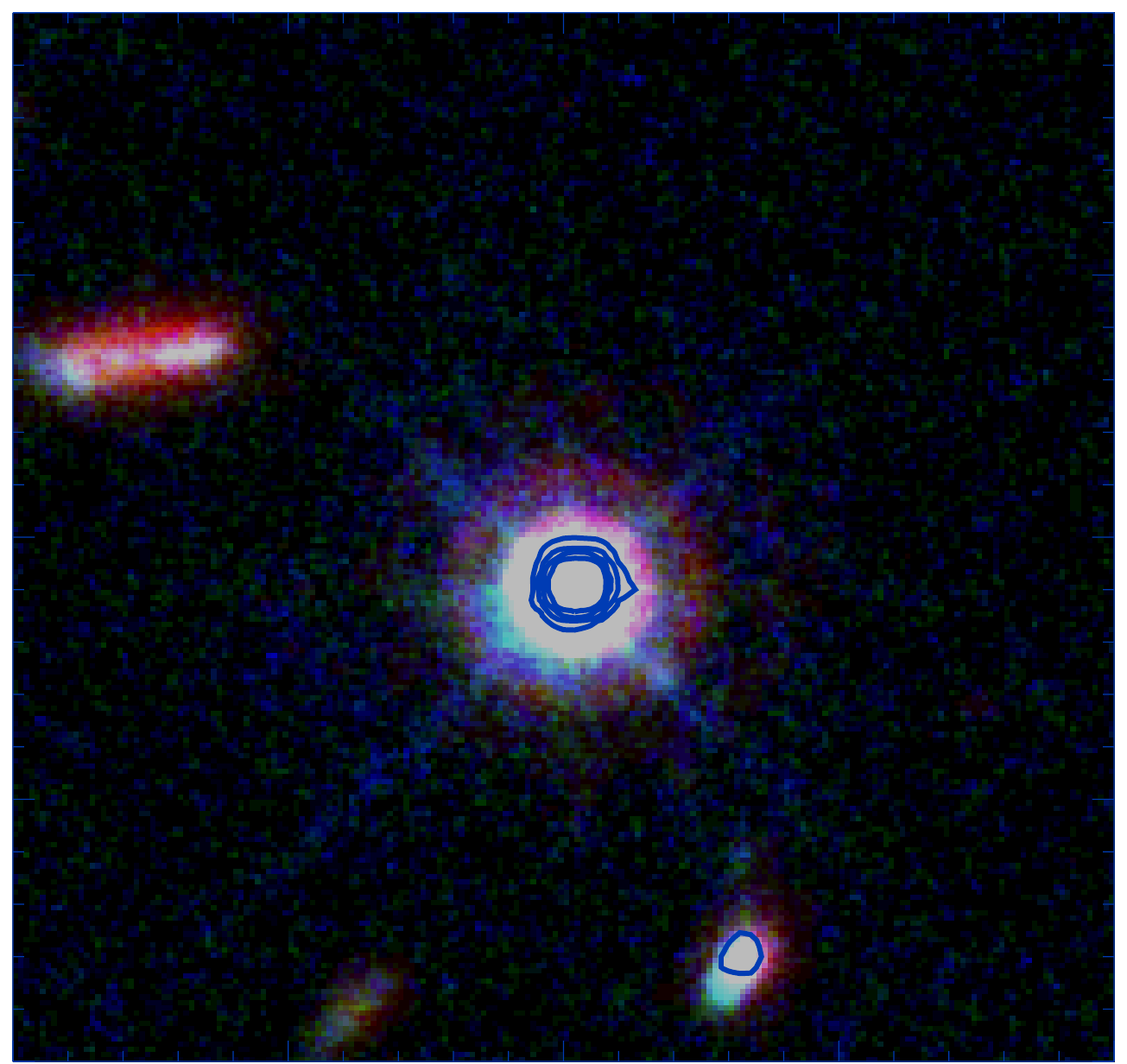}\\
  
  \begin{overpic}[trim=140 230 92 197,clip,width=2.6in,height=2.6in]{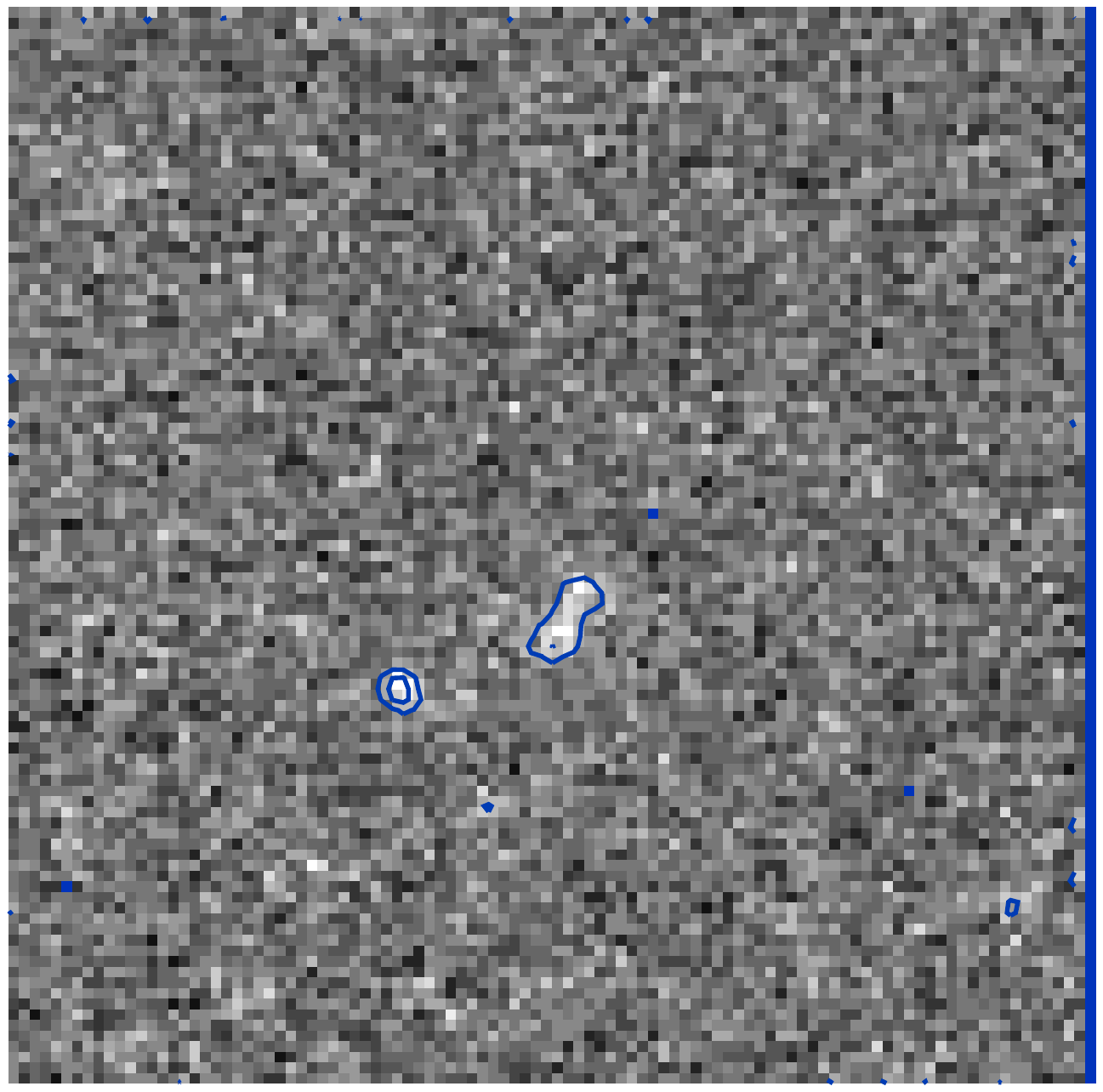} \put(7,87){\large \color{white} \textbf{GN-UVC-2}}\end{overpic} &

  \includegraphics[trim=118 247 118 190,clip,width=2.6in,height=2.6in]{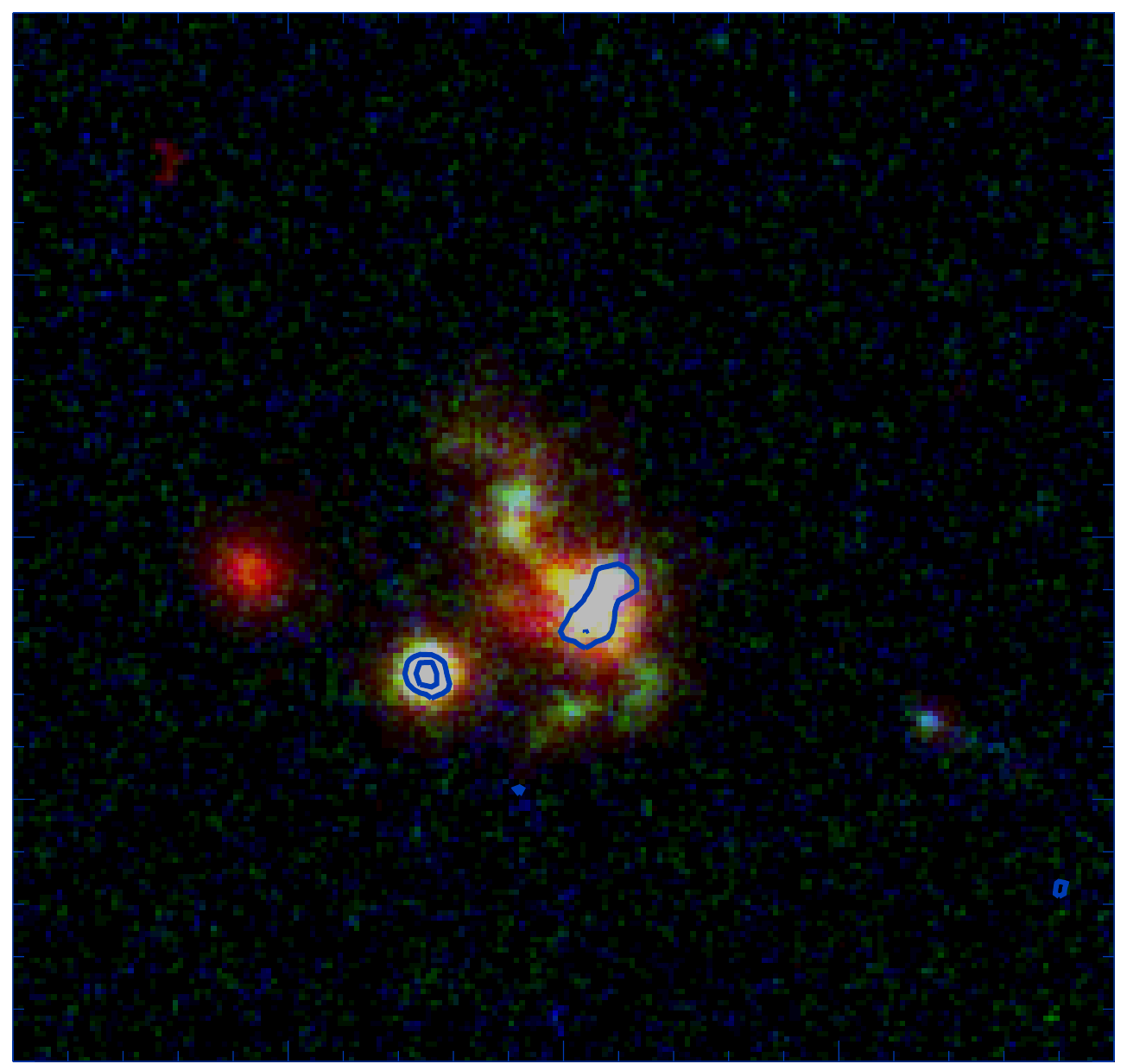}\\
  
  \begin{overpic}[trim=140 230 92 197,clip,width=2.6in,height=2.6in]{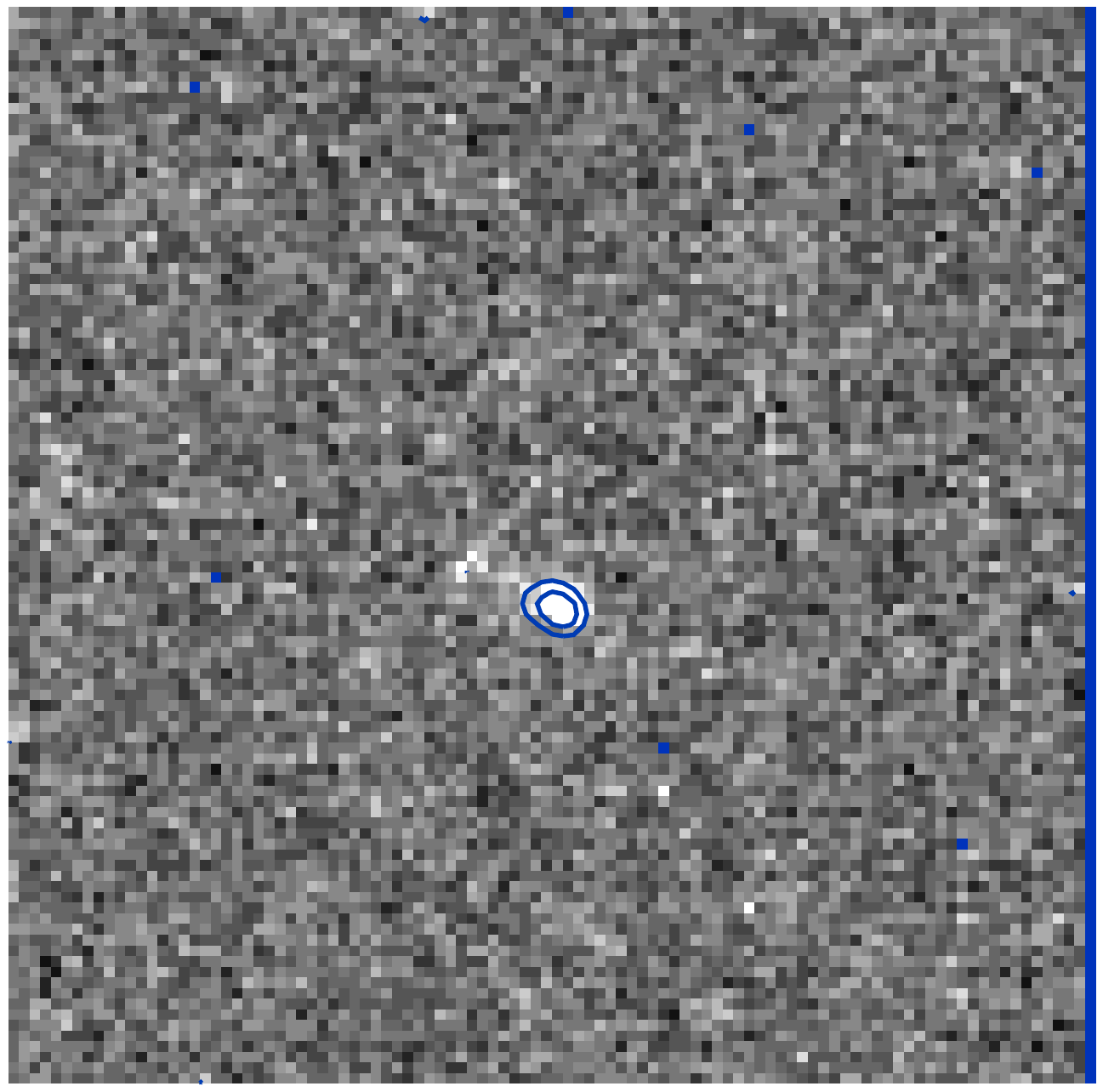} \put(7,87){\large \color{white} \textbf{GN-UVC-3}}\end{overpic} &

  \includegraphics[trim=118 247 118 190,clip,width=2.6in,height=2.6in]{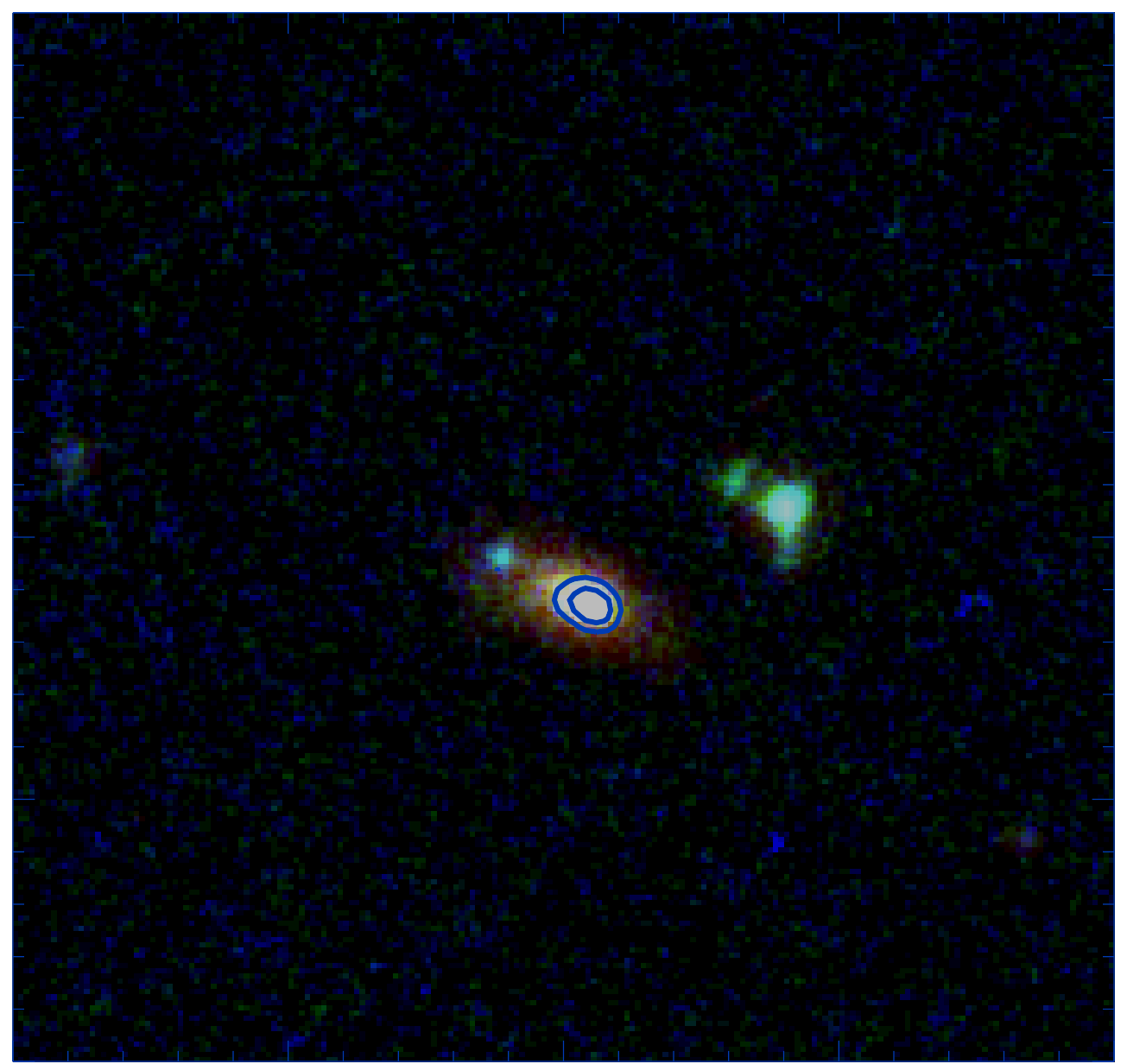}\\

\end{tabular}
\caption{
F275W thumbnails (left) and three-color images (red = F160W, green = F606W, and blue = F435W) of our six candidate LyC emitters. Blue contours show F275W emission for sources detected at or above the 4$\sigma$ level in the HDUV F275W imaging. (Note that GN-UVC-6 is detected at the 2.9$\sigma$ level in F275W and was selected based on its relatively blue F275W-F435W color; see Section \ref{results}.) Images are $6''$ on a side. North is up and East is to the left. The sources appear slightly below center in $y$ to allow for the labels at the top.
}
\label{pascal_thumbs}
\end{figure*}  
\begin{figure*}
\setcounter{figure}{3}
\centering
\begin{tabular}{p{2.62in}p{2.62in}}
  
  \begin{overpic}[trim=140 230 92 197,clip,width=2.6in,height=2.6in]{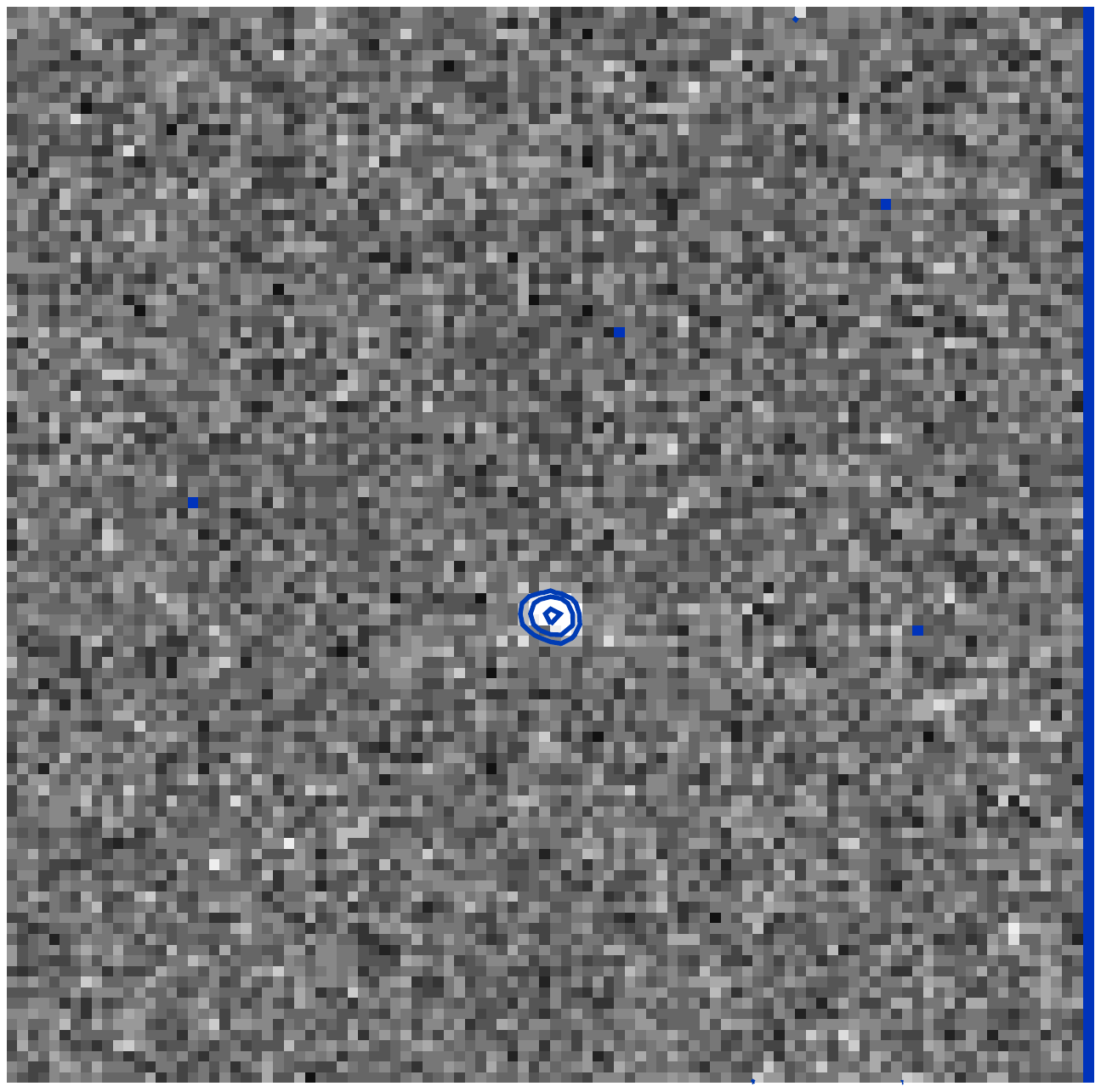} \put(7,87){\large \color{white} \textbf{GN-UVC-4}}\end{overpic} &

  \includegraphics[trim=118 247 118 190,clip,width=2.6in,height=2.6in]{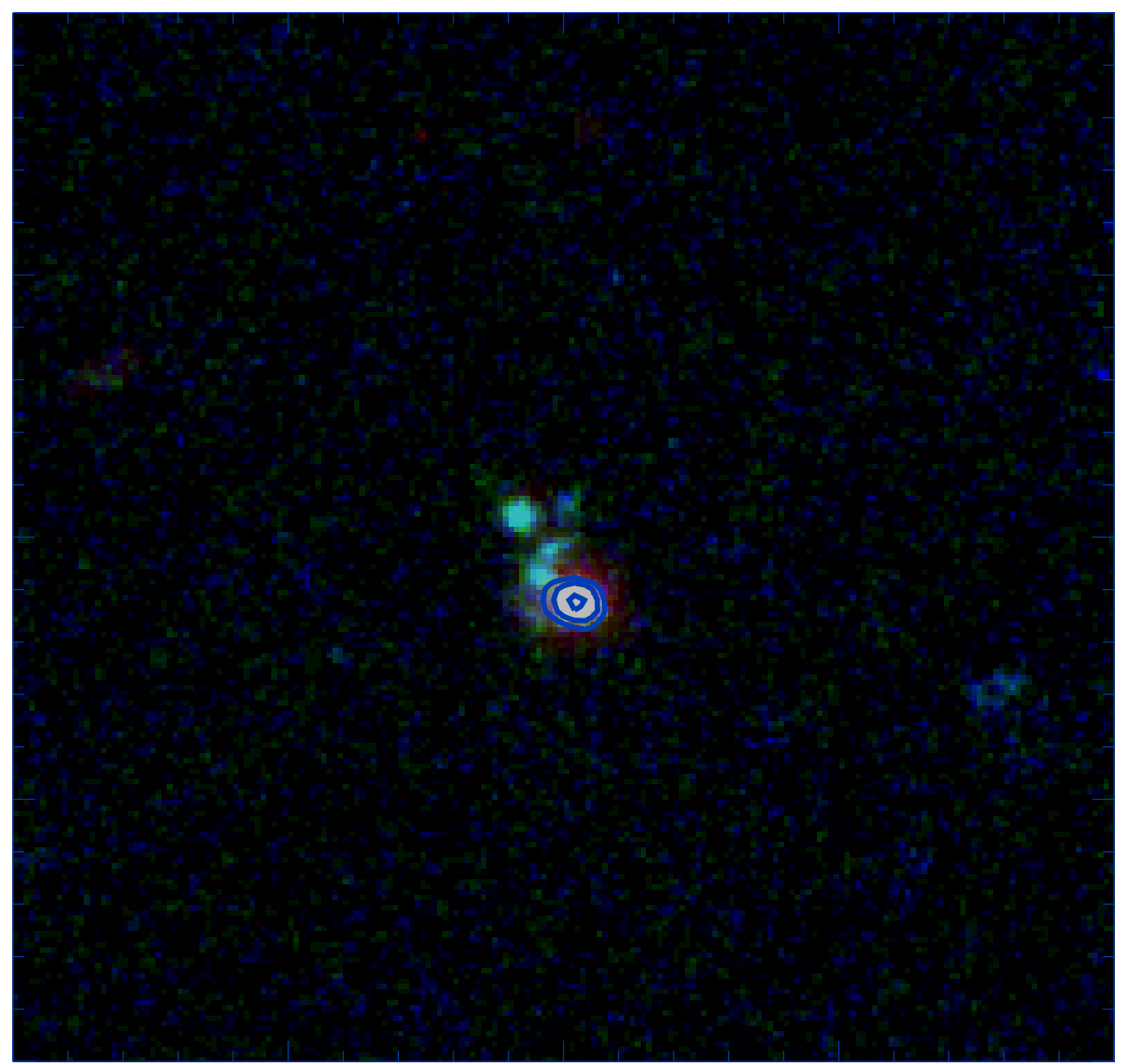}\\
  
  \begin{overpic}[trim=140 230 92 197,clip,width=2.6in,height=2.6in]{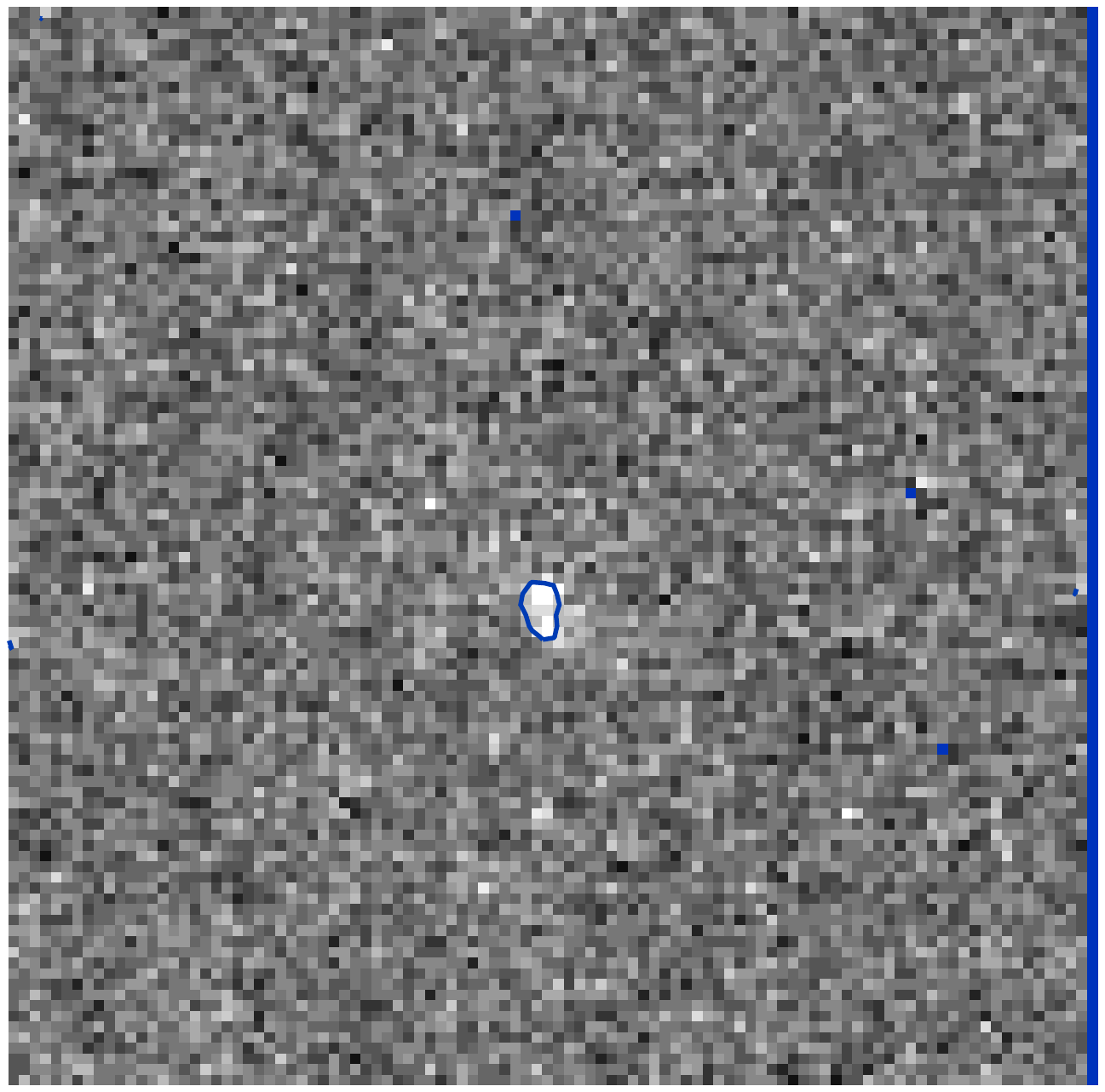} \put(7,87){\large \color{white} \textbf{GN-UVC-5}}\end{overpic} &

  \includegraphics[trim=118 247 118 190,clip,width=2.6in,height=2.6in]{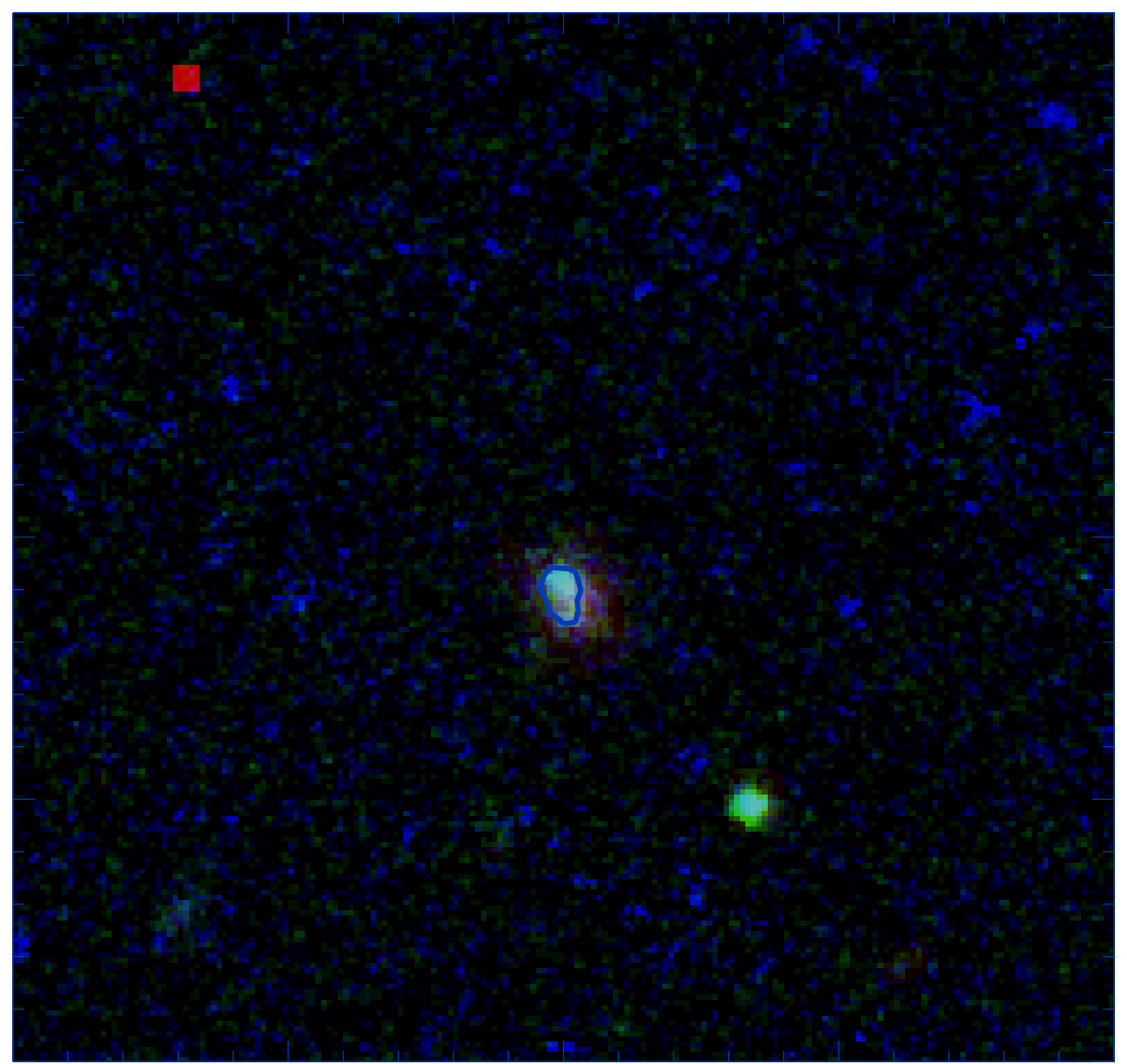}\\
  
  \begin{overpic}[trim=140 230 92 197,clip,width=2.6in,height=2.6in]{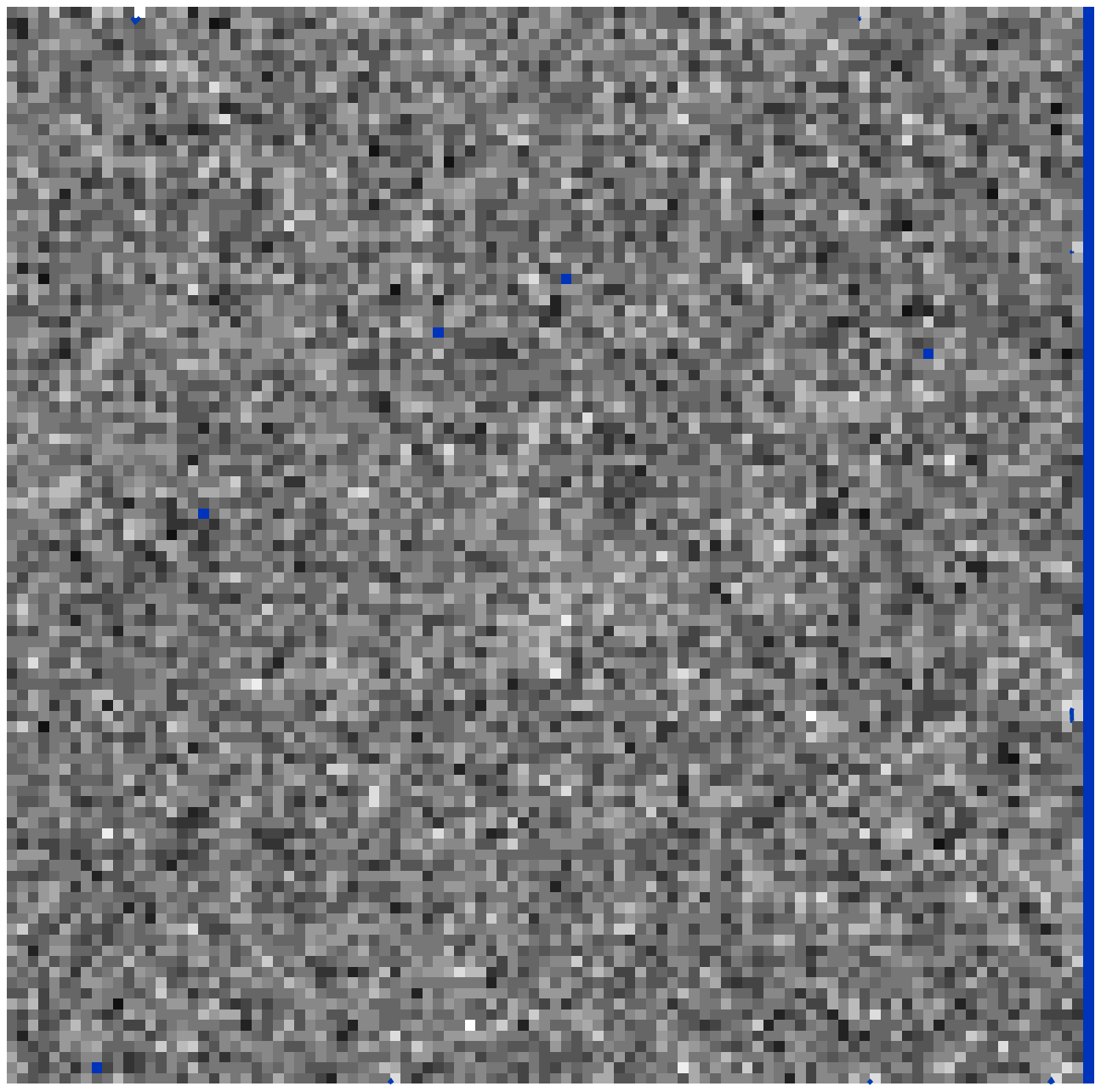} \put(7,87){\large \color{white} \textbf{GN-UVC-6}}\end{overpic} &

  \includegraphics[trim=118 247 118 190,clip,width=2.6in,height=2.6in]{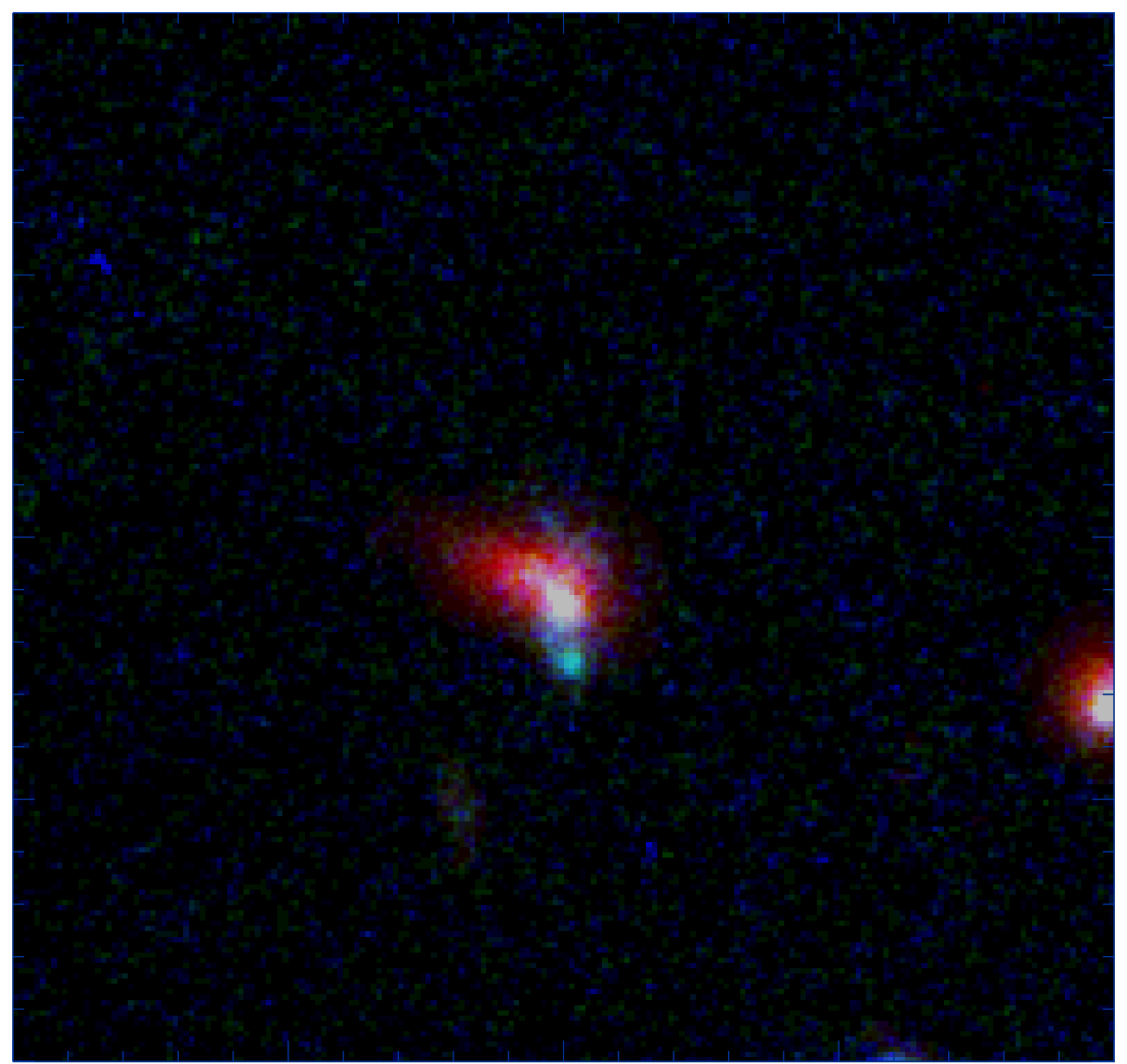}\\
\end{tabular}
\caption{
(Cont.)
}
\end{figure*}

\subsubsection{GN-UVC-1}
\label{gnuvc1}
The broad-line quasar GN-UVC-1 at $z = 2.583$ with an F275W magnitude of 
23.14 is easily the brightest of our six candidate LyC emitters. 
The smaller F275W source to its lower right (see Figure~\ref{pascal_thumbs}) 
is likely a star-forming galaxy at low redshift. 
GN-UVC-1 is one of two objects in our candidate sample (the other being GN-UVC-3) 
that was selected both by its F275W flux alone and by its relatively blue F275W-F435W 
color ($\approx 2.6$ mag). As shown in Figure \ref{fig_quasar}, the \hst{}/WFC3 G280 
grism spectrum of GN-UVC-1 (GO12479, PI: Hu) directly confirms its identification as 
a high-redshift LyC emitter.

\begin{figure}[t]
\setcounter{figure}{4}
\includegraphics[trim=12 0 12 0,clip,width=3.35in,angle=0]{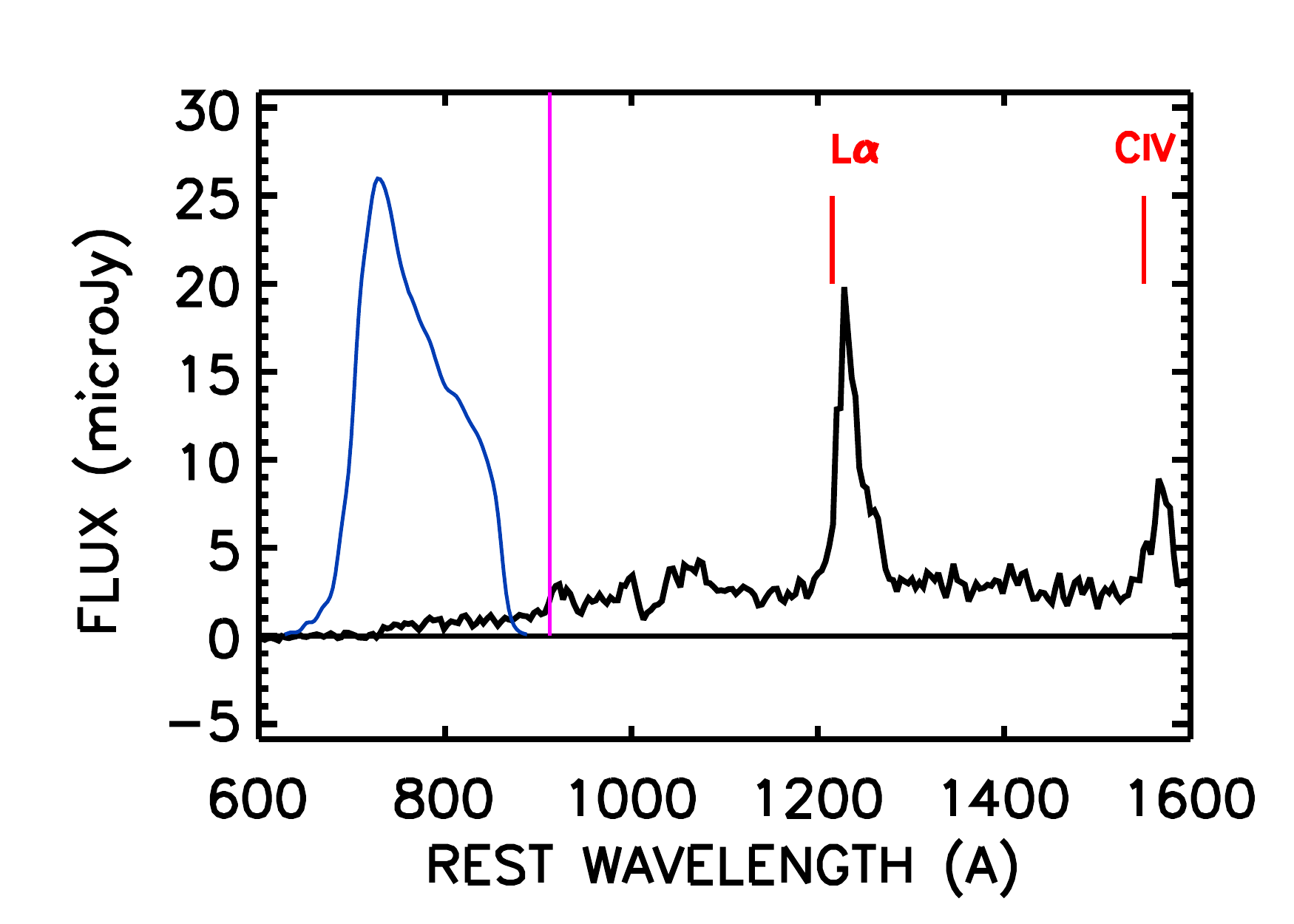}
\caption{
G280 grism spectrum from \hst{}/WFC3 for GN-UVC-1. The blue curve shows the relative response of the F275W filter shifted into the rest frame of GN-UVC-1, and the pink vertical line marks the LyC edge.
\label{fig_quasar}
}
\end{figure}

\subsubsection{GN-UVC-2}
\label{gnuvc2}
\begin{figure}[t]
\includegraphics[trim=12 0 0 0,clip,width=3.35in,angle=0]{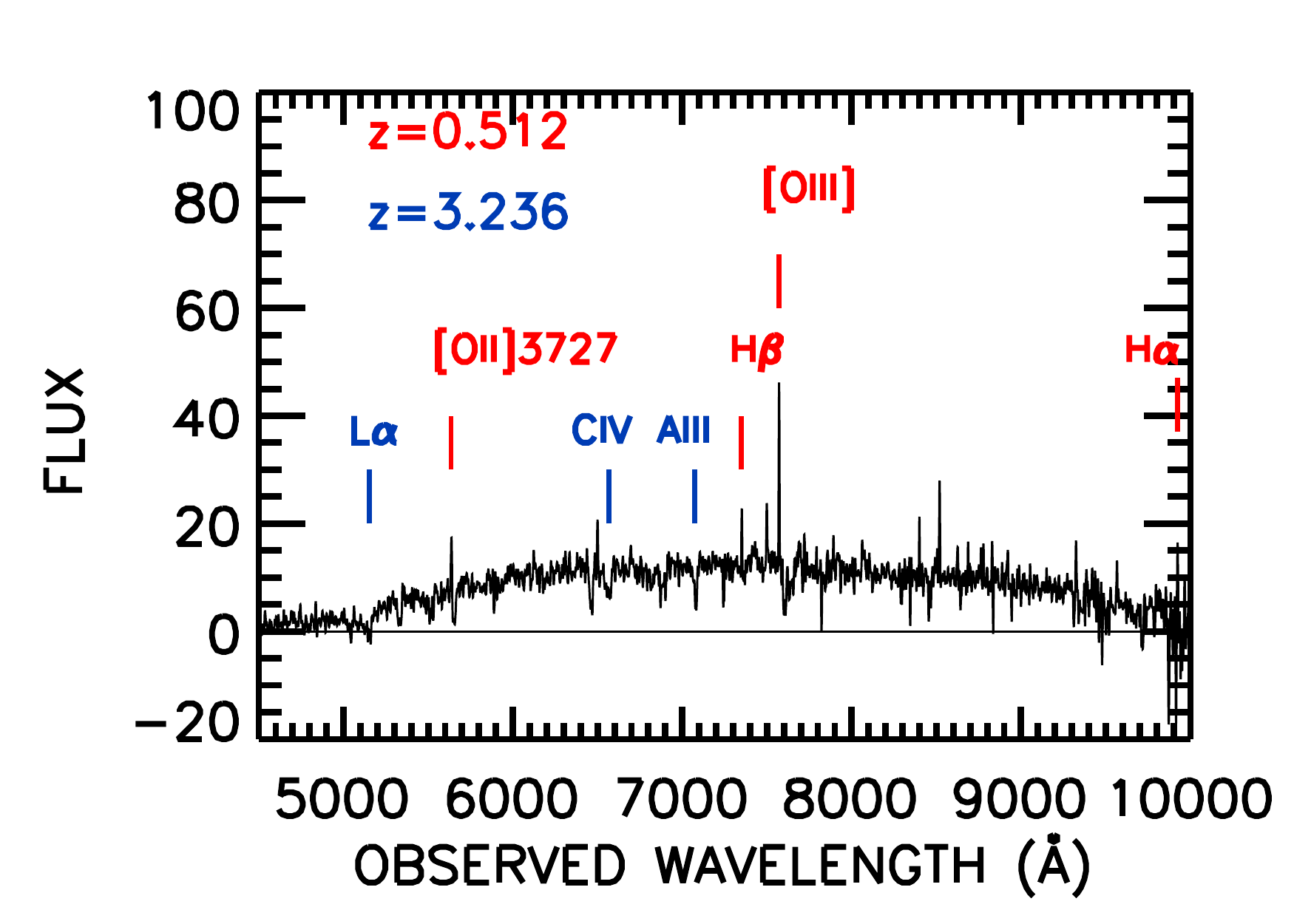}
\caption{
Our DEIMOS spectrum of GN-UVC-2. In addition to absorption features from a $z$ = 3.236 galaxy (marked in blue), emission lines from a foreground galaxy at $z$ = 0.512 (marked in red) are clearly present.
\label{pascal2}
}
\end{figure}

GN-UVC-2 illustrates particularly well the difficulties of trying to confirm LyC emission from
high-redshift galaxies. 
There are two positions in the F275W image (see Figure~\ref{pascal_thumbs}) that show significant UV flux: one coinciding with a somewhat extended star-forming galaxy/possible weak AGN ($L_{X} \sim 6\times10^{42}$ erg s$^{-1}$) roughly at image center, and one coinciding with a neighboring source about 1$\arcsec$ away. 
In Figure~\ref{pascal2}, we show our DEIMOS spectrum with a total exposure time of $\sim$6 hours. In the individual exposures, we used a 1$\arcsec$ wide slit and slit position angles ranging from 41$^{\circ}$ to 59$^{\circ}$. 
We visually identify two redshift systems in the spectrum.
Absorption features from the extended, central $z = 3.236$ source 
(L$\alpha$, CIV1550, and AlIII1670, marked in blue on
the spectrum) are clearly present, but so are emission lines
([OII]3727, H$\beta$ and [OIII]4959,5007) 
from a $z = 0.512$ foreground source (marked in red). 
We note that at the position angles of the individual spectra, the neighboring source is located just outside the slit. Thus, it is unlikely that it could be the source of the emission lines, since the lines would have to be extraordinarily strong to overflow into the slit at these position angles. Moreover, the emission lines seen in the individual spectra are invariant from exposure to exposure, despite changes in position angle. This suggests that the low-redshift emission lines come from a source superposed more or less directly on top of the $z = 3.236$ galaxy, calling into question the origin of the measured F275W flux.

\begin{figure}[t]
\includegraphics[width=3.3in,angle=0]{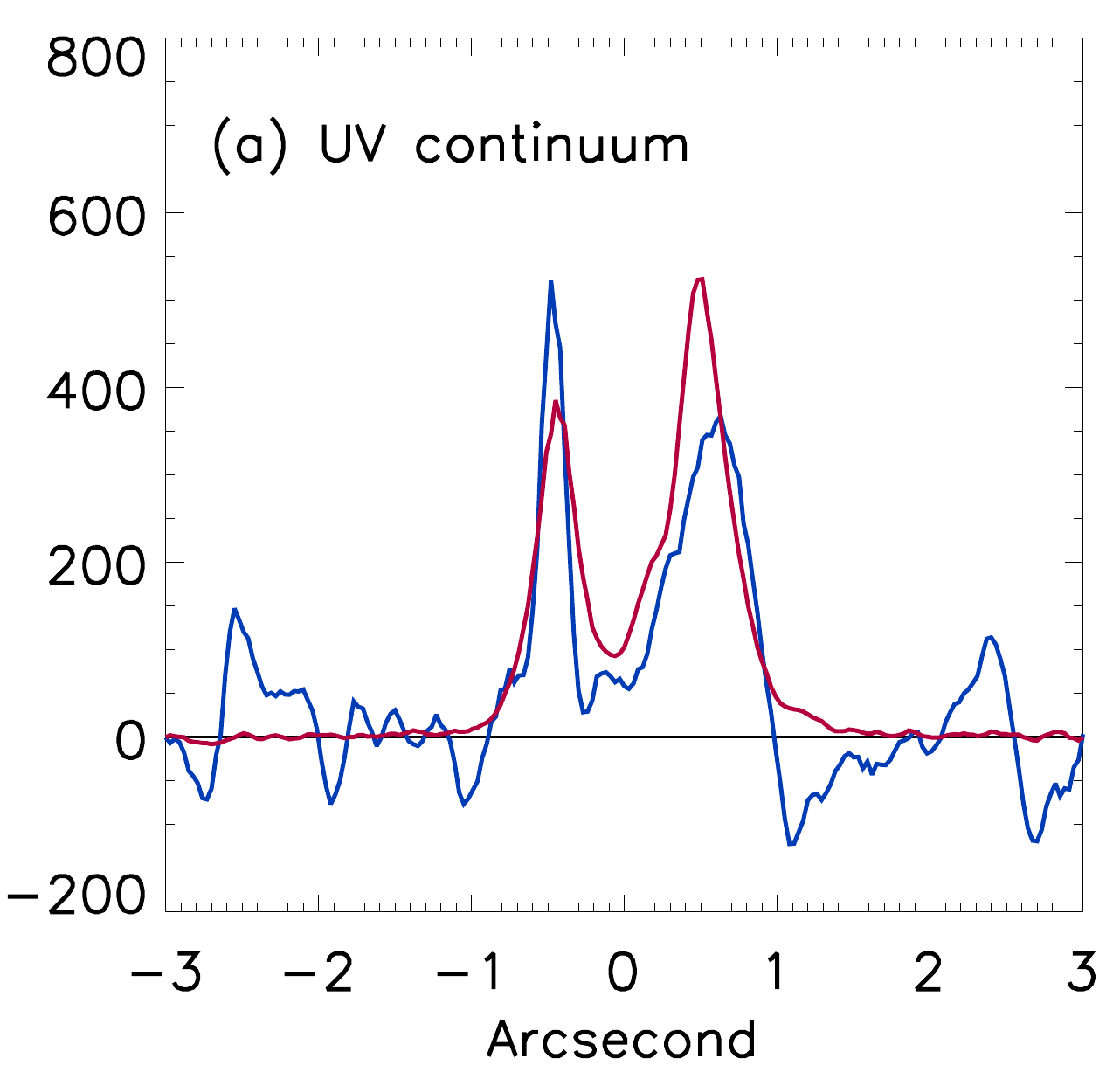}
\includegraphics[width=3.3in,angle=0]{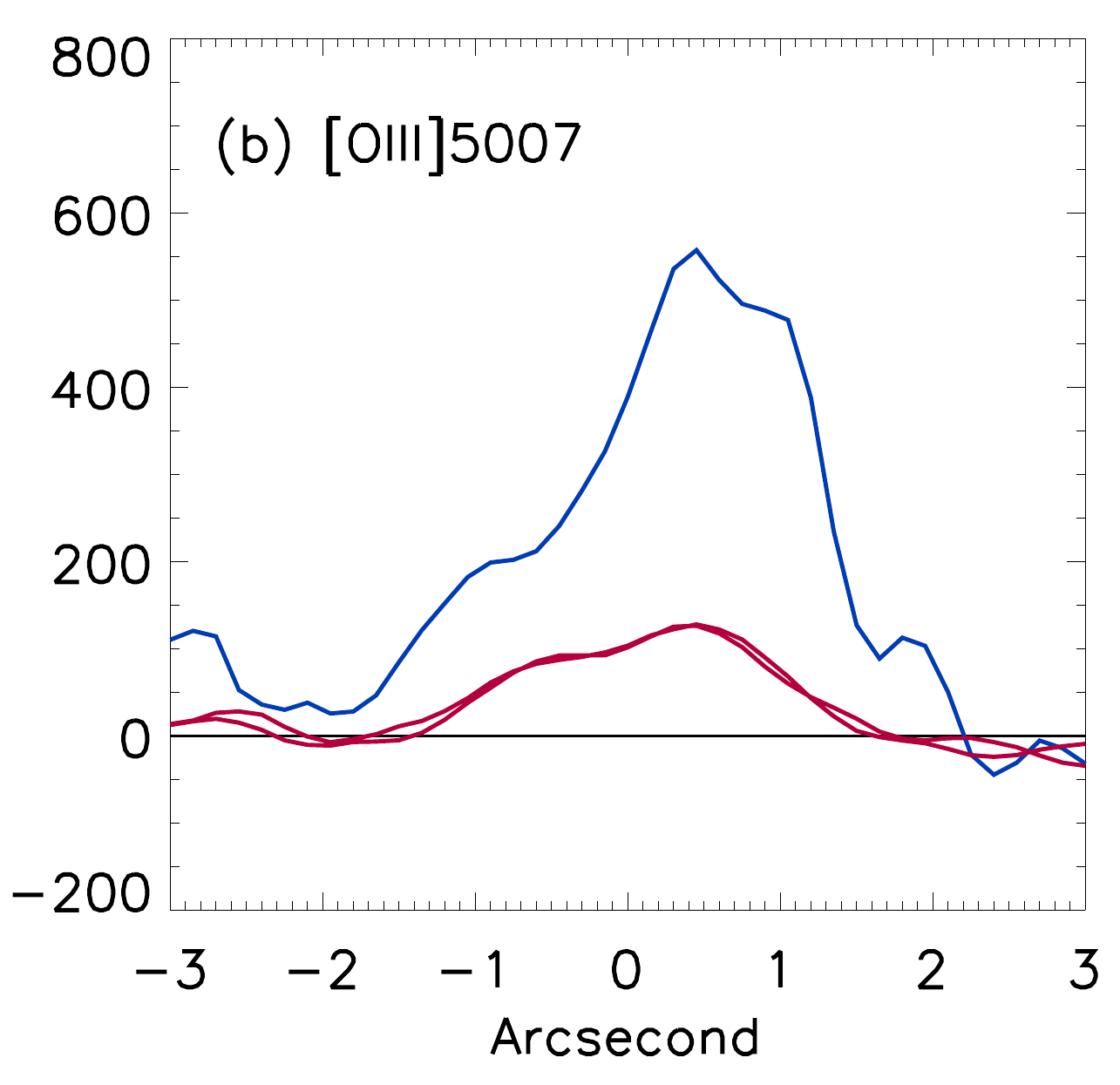}
\caption{
(a) The F275W (blue curve) and F160W (red curve) light profiles as they would appear in a $1\arcsec$ wide slit at a position angle of 116$^{\circ}$ that crosses both the LBG and the neighboring source in the GN-UVC-2 image (upper-right thumbnail in Figure~\ref{pascal_thumbs}). The relative normalization of the two profiles in this panel is arbitrary. 
(b) The light profile of the $z$ = 0.512 [OIII] $\lambda$5007 \AA~line (blue curve) and the continuum measured both 300 \AA ~redward and blueward of the line (red curves) as seen in a 1~hour Keck/DEIMOS spectrum taken at a position angle of 116$^{\circ}$. The 0\farcs6 seeing smooths the profile considerably relative to the \hst{} continuum data, but the [OIII] profile lies above the continuum throughout the profile and is clearly present at the positions of both the LBG and the neighboring source.
\label{pascal_profile_fig}
}
\vspace{0.1cm}
\end{figure}

In Figure~\ref{pascal_profile_fig}(a), we show the F275W (blue curve) and F160W (red curve) continuum light profiles as they would appear in a 1$\arcsec$ wide slit at a position angle of 116$^{\circ}$ that covers both the central and neighboring sources (the relative normalization of the profiles is arbitrary). There is significant UV continuum flux at both positions. 

We next obtained an additional 1~hour DEIMOS spectrum (1$\arcsec$ wide slit and 0\farcs6 seeing) at this position angle.  If we examine the [OIII] $\lambda$5007 light profile from this new spectrum (Figure~\ref{pascal_profile_fig}(b)), we see that there is [OIII] emission (blue curve) at both positions. The [OIII] profile is somewhat smoothed relative to the \hst{} continuum data in Figure \ref{pascal_profile_fig}(a) due to the seeing.  However, at the positions of both the central and neighboring sources, it is significantly brighter than the continuum measured both 300~\AA ~redward and blueward of the line center (red curves). This confirms that the neighboring source is also at $z \sim$ 0.5. 

We consequently interpret the low-redshift emission lines as coming from a $z \sim$ 0.5 galaxy with two spatially separated star-forming components (i.e., similar to the chain galaxies of \citet{cow95} and references therein), one of which lies directly along the line of sight to the high-redshift Lyman-break galaxy (LBG). Thus, the bulk of the measured F275W flux probably comes from the low-redshift galaxy, meaning GN-UVC-2 should not be used when constraining the ionizing background at $z\sim 3$. We exclude it from our analysis in Section~\ref{flux}.

\begin{figure}[t]
\includegraphics[trim=12 0 0 0,clip,width=3.35in,angle=0]{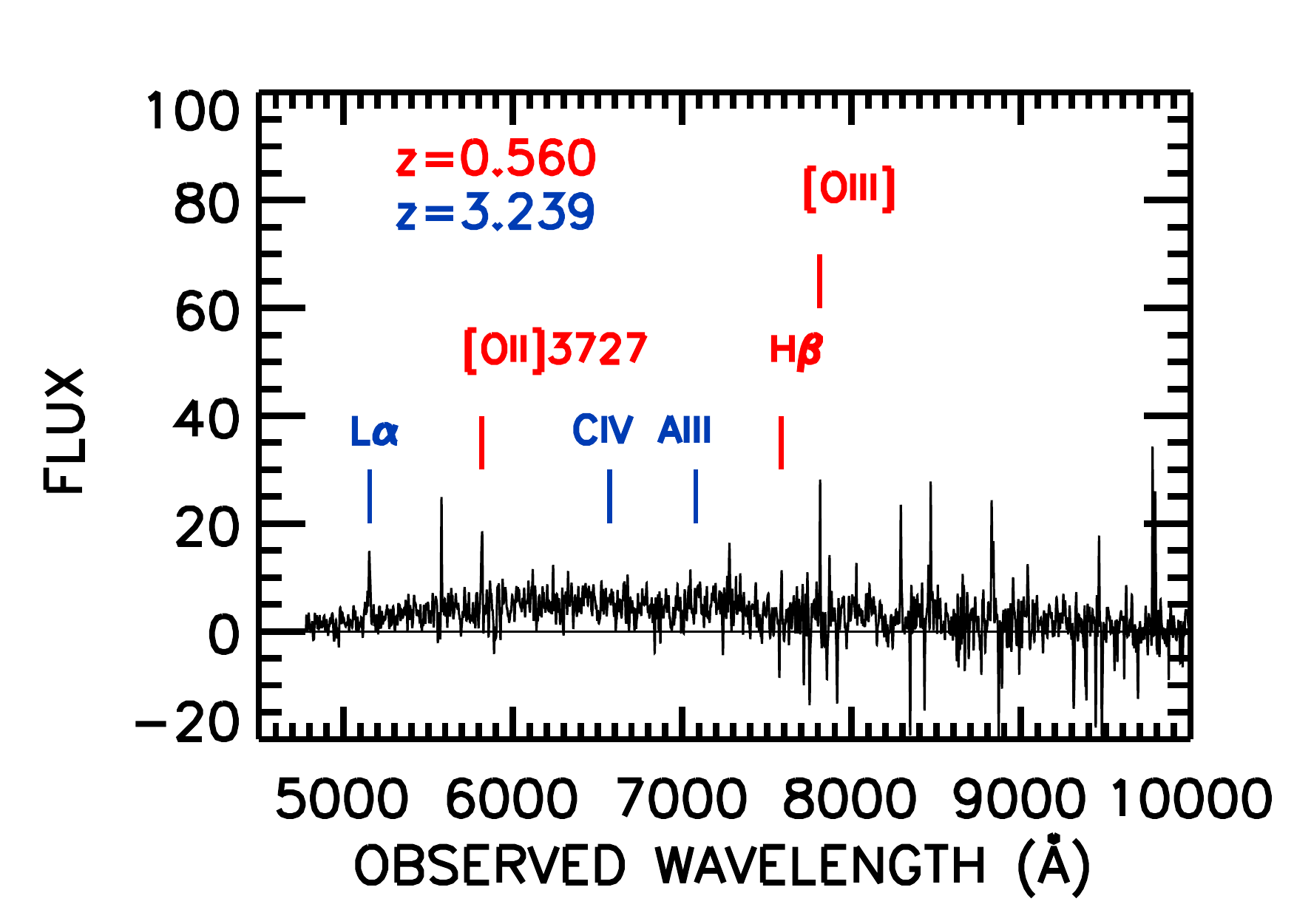}
\caption{
Our DEIMOS spectrum of GN-UVC-3, which shows features from sources at $z = 3.239$ (blue)
and $z = 0.56$ (red).
\label{uvc3spec}
}
\end{figure}

\subsubsection{GN-UVC-3}
\label{gnuvc3}
We show in Figure~\ref{uvc3spec} our DEIMOS spectrum of GN-UVC-3. 
As with GN-UVC-2, 
spectral features from a $z = 3.239$ source (Ly$\alpha$ emission and CIV1550 
absorption marked in blue) and from a foreground $z = 0.56$ source
 ([OII]3727, H$\beta$, and [OIII]4959,5007 emission marked in red) 
are visually identified. The high-redshift system was previously
identified by \citet{red06}. The low-redshift interloper is almost certainly the source of the F275W flux. 
Thus, we also exclude GN-UVC-3 from our analysis in Section~\ref{flux}.

\subsubsection{GN-UVC-4}
\label{gnuvc4}
\begin{figure}[t]
\includegraphics[trim=12 0 0 0,clip,width=3.35in,angle=0]{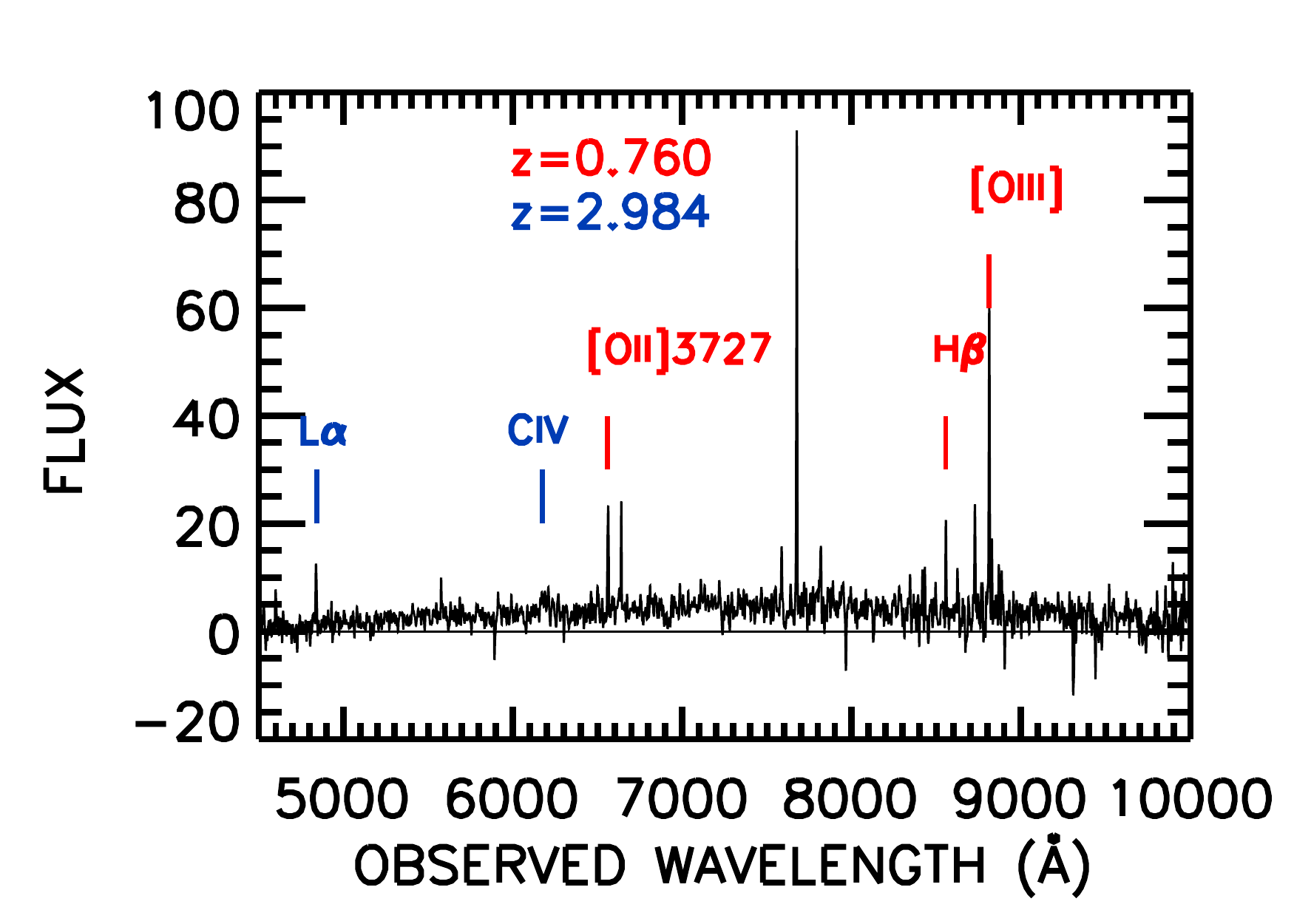}
\caption{
Our DEIMOS spectrum of GN-UVC-4, which shows features from sources at $z = 2.984$ (blue) 
and $z = 0.76$ (red).
\label{gnuvc4spec}
}
\end{figure}

GN-UVC-4 is peculiar, because its \textit{BVH} thumbnail in 
Figure~\ref{pascal_thumbs} shows at least two differently-colored components
(a redder source at image center and an elongated, clumpy, blue source extending northward). However, the UV emission corresponds only to the central redder source.
Our DEIMOS spectrum confirms the source as a projection of two emission
line galaxies at 
very different redshifts (see Figure \ref{gnuvc4spec}). 
The high-redshift identification
at $z=2.984$ is based on Ly$\alpha$ and CIV1550 emission and confirms the redshift
obtained by \citet{red06}. The low-redshift identification at $z=0.760$ is based on
[OII]3727, H$\beta$, and [OIII]4959,5007 emission.  We conclude that the UV emission
probably comes from the low-redshift galaxy.
Thus, we exclude GN-UVC-4 from our analysis in Section \ref{flux}.

\subsubsection{GN-UVC-5}
\label{gnuvc5}
\begin{figure}[t]
\includegraphics[trim=12 0 0 0,clip,width=3.35in,angle=0]{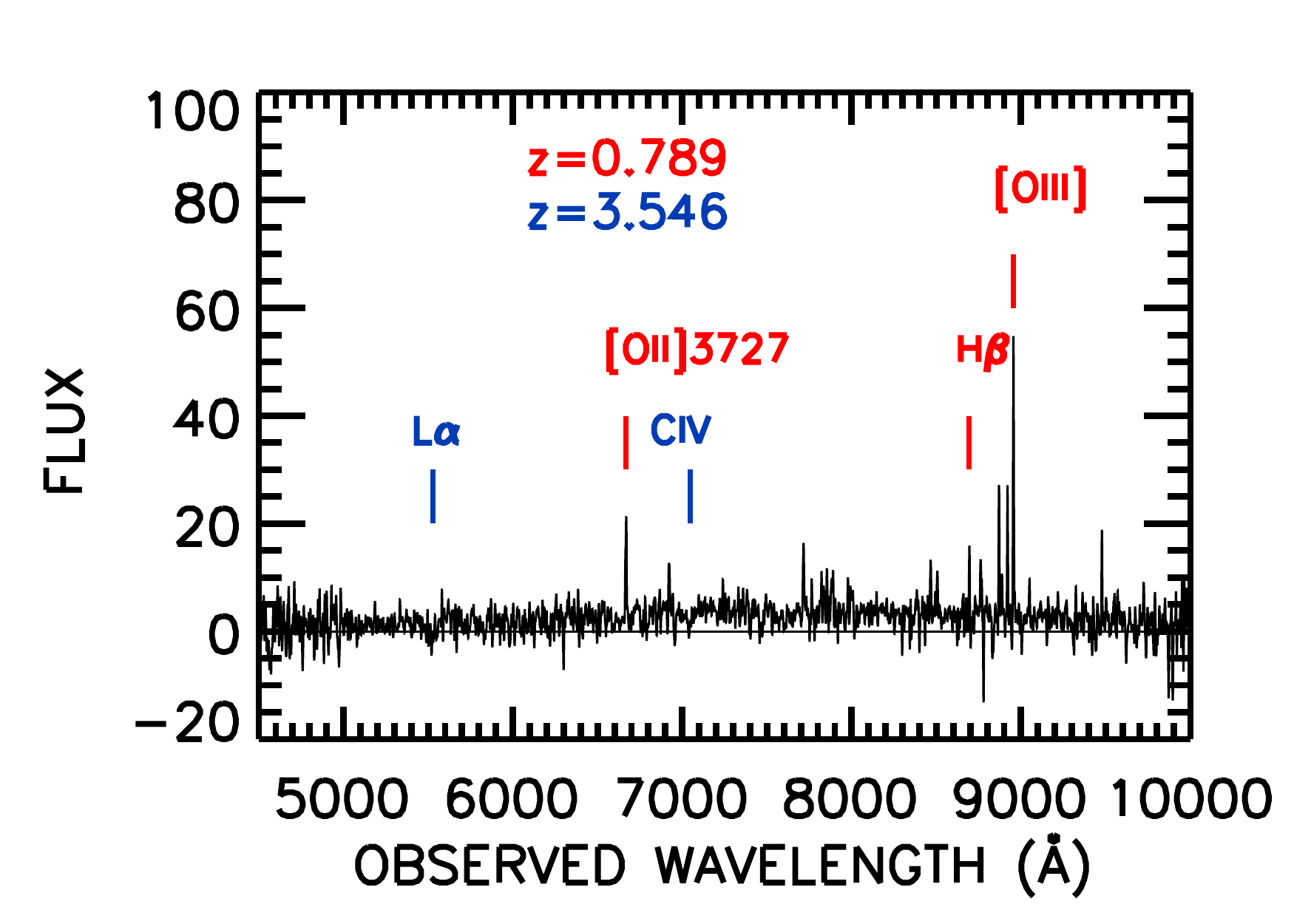}
\caption{
Our DEIMOS spectrum of GN-UVC-5, which shows features from sources at 
$z = 3.546$ (blue) and $z = 0.789$ (red).
\label{gnuvc5spec}
}
\end{figure}

The F275W detection of GN-UVC-5 is quite surprising, because
by $z = 3.546$, the F275W bandpass probes rest-frame wavelengths well 
below the LyC break (at $\sim$590 \AA), where we expect virtually no transmission 
of ionizing radiation from the galaxy due to attenuation by the IGM (e.g., \citealt{in14}).
\citet{u15} label this object as having a very secure redshift identification 
(quality code `A'), while photometric redshift estimates given 
in the 3D-HST catalog \citep{3dhst} put GN-UVC-5 (their GN-26359) at 
$z_{phot} = 0.74$. These conflicting redshift estimates, together with the 
apparently singular nature in F275W of GN-UVC-5 (see Figure~\ref{pascal_thumbs}), 
suggest that this source may be yet another chance projection of two galaxies 
at vastly different redshifts. 
Indeed, our DEIMOS spectrum shows weak Ly$\alpha$ emission and
CIV1550 in absorption (see Figure \ref{gnuvc5spec}), confirming the redshift of 
U et al.\ (2015), while also showing [OII]3727, H$\beta$, and [OIII]4959,5007 
emission from a low-redshift galaxy at $z = 0.789$.
Since the UV emission probably comes from the low-redshift galaxy,
we exclude GN-UVC-5 from our analysis in Section~\ref{flux}.

\subsubsection{GN-UVC-6}
\label{gnuvc6}
Of our six candidate LyC emitters, this is the sole object selected only 
based on its relatively blue F275W-F435W color ($\approx 1.6$ mag, bluer even than the bright quasar GN-UVC-1), though we note that it is detected at the 2.9$\sigma$ level in F275W. We find a possible far-infrared counterpart to this source in the GOODS-$Herschel$ catalog of \citet{elbaz11} (separation $< 0.5$\arcsec). This may indicate the presence of an AGN, though with only one detection from $Herschel$ (in the PACS 160 $\mu$m band) and negligible X-ray flux, this is somewhat tentative. The complicated morphology and multiple \textit{BVH} colors seen in GN-UVC-6's three-color thumbnail (see Figure \ref{pascal_thumbs}) 
suggests there may be superposed sources that could lie at different redshifts. However, the photometric redshift of $z=2.38$ from \citet{raff11} is in good agreement with the spectroscopic redshift of $z = 2.439$ from \citet{red06}, which argues against contamination from foreground objects. We recently obtained a DEIMOS spectrum of this source, which confirms that there are no emission features in the $4500 - 10,000$ \AA\ range that would indicate the presence of a superposed foreground object. We conclude that GN-UVC-6 remains a good candidate LyC emitter and can be used to obtain limits on the contribution of galaxies 
to the ionizing background at $z \sim 3$.

\section{Contributions to the Ionizing Flux}
\label{flux}
Determining the absolute escape fraction for each of our candidate LyC emitters is challenging, if not impossible, without knowledge of each source's intrinsic spectral energy distribution (SED) and degree of reddening. Instead, we compute the ionization fraction from $f_{ion} = f_{LyC}/f_{1500}$; that is, the ratio of the flux at the rest-frame LyC wavelength ($\sim$675 \AA~at $z = 3$)
to the flux at rest-frame 1500~\AA, ignoring the small differential $K$-correction as a function of redshift. We use the F606W flux as a rough estimate of the rest-frame 1500~\AA~flux, assuming a flat $f_{\nu}$ SED. We note that this approximation is most accurate for sources very near $z = 3$. We give our measured ionization fractions in the last column of Table \ref{sources} for the two sources that do not have clear spectroscopic evidence for contamination by foreground galaxies.

\subsection{Quasar UV Emissivity}
\label{quasarUV}
GN-UVC-1 is brighter than any other candidate in our sample by approximately two magnitudes. Since it is also the only quasar, we consider its contribution to the ionizing background separately from our other candidates. We measured its flux density at 912~\AA~directly from its UV grism spectrum after renormalizing the spectrum to match the total flux detected in the F275W imaging data. We then converted this to an ionizing volume emissivity, $\epsilon_{912}$, defined as the luminosity density per unit frequency divided by the comoving volume over the redshift range $z=2.439-3.546$ (the lowest and highest redshifts of our candidates).
We hereafter quote measurements and uncertainties of $\epsilon_{912}$ in units of $10^{24}$~erg~s$^{-1}$~Hz$^{-1}$~Mpc$^{-3}$. 
For the single source in our ``quasar sample", the Poisson noise dominates; from \citet{gehrels86}, the 68\% confidence range for one object is 0.173 to 3.300. We find a quasar UV emissivity, $\epsilon_{912,q}$, of $8.3^{27}_{1.4}$.

\begin{figure}[t]
\includegraphics[trim=30 10 60 10,clip,width=3.35in,angle=0]{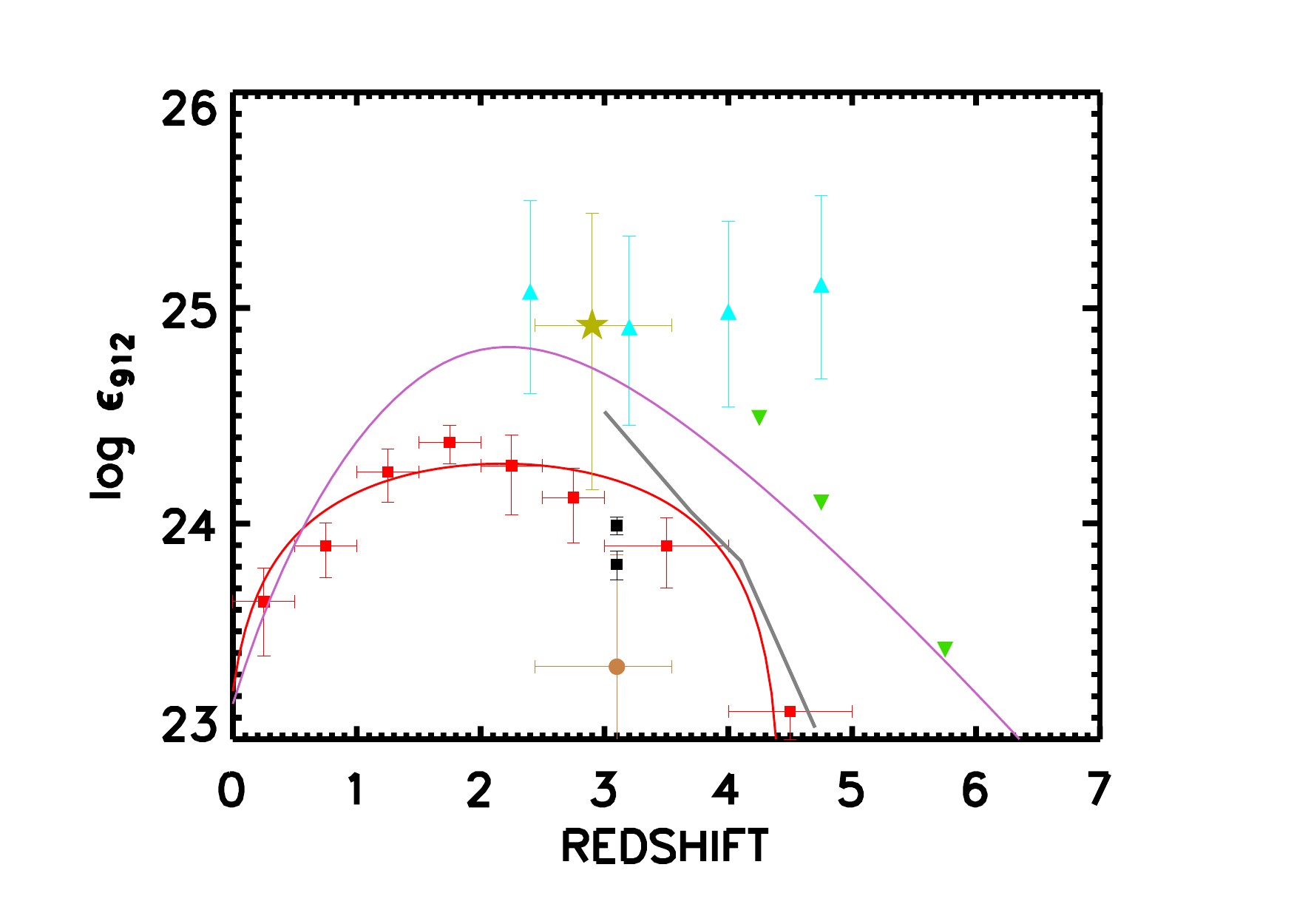}
\caption{Ionizing volume emissivity at $z = 3$ estimated from the quasar GN-UVC-1 (gold star) and compared to literature results for quasar/AGN contributions to the ionizing background at similar redshifts. Red, green, and black symbols show data from CBT09, \citet{parsa18}, and \citet{mich16}, respectively, and the red, purple, and grey curves are from CBT09, \citet{hm12}, and \citet{meik05}, respectively. The bronze circle shows the ionizing emissivity from star-forming galaxies and low-luminosity AGNs identified in this work (e.g., GN-UVC-6). Horizontal error bars on our data points reflect the range of redshifts in our candidate sample; the symbols have been splayed around the mean redshift of 3 for clarity. Cyan triangles show the observed ionizing emissivity from \citet{beck13}.
\label{literature}
}
\end{figure}

\subsection{Non-quasar UV Emissivity}
\label{nonquasarUV}
With four of five non-quasar candidates showing clear contamination from foreground objects, our ``star-forming galaxy" sample considered here consists of GN-UVC-6 only. We estimated an ionizing emissivity for star-forming galaxies and low-luminosity AGNs, $\epsilon_{912,g}$, by assuming that GN-UVC-6's F275W flux is entirely at the filter's effective wavelength of $\sim 2704$~\AA. This wavelength probes the LyC at $\sim786$ \AA~in the rest frame of GN-UVC-6. To allow for a simpler and more direct comparison with literature results, which mostly consider the ionizing volume emissivity at or near the LyC edge, we scaled the measured flux density to that at 912~\AA~following the results of \citet{lusso15}. They used a sample of 53 quasars at $z \sim 2.4$ to construct a stacked UV spectrum between 600 and 2500~\AA~(rest frame), correcting for both intergalactic Lyman forest and Lyman continuum absorption, and found a $\lambda < 912$~\AA~continuum slope of $\alpha_{\nu} = -1.70$. We used this power law slope to do our scaling (e.g., from $\sim786$~\AA~to 912~\AA~for GN-UVC-6). We find that star-forming galaxies and low-luminosity AGNs contribute $\epsilon_{912,g} = 0.22 ^{0.72}_{0.04}$ to the ionizing background at $z\sim3$, where the total error is again dominated by the Poisson noise (in the 68\% confidence range).

\subsection{Comparisons with the Literature}
\label{emissivity}

In Figure \ref{literature}, we put our measurements in the context of other $z\sim 3$ measurements from the literature. The level of ionizing volume emissivity that we estimated from our single quasar (gold star) is a factor $\sim$38 larger than our $\epsilon_{912,g}$ (bronze circle). Though our small sample size makes quantitative comparisons difficult, our
$\epsilon_{912,q}$ is consistent, within the very large uncertainties, with the contribution measured by CBT09 from their much larger sample of broad-line AGNs (their Equation~1; red curve and points in Figure~\ref{literature}). It is also roughly consistent with quasar ionizing emissivity results from \citet{meik05} (grey curve) and \citet{hm12} (purple curve). 

\citet{beck13} used Ly$\alpha$ forest observations to infer the total ionizing background from $2 < z < 5$. They obtained a nominal $\epsilon_{912}$ = 8.15 at $z = 3.2$, again a factor of $\sim 38$ larger than our upper-limit estimate of $\epsilon_{912,g}$ but consistent with our measured contribution from quasars. This suggests that quasars alone contribute virtually all of the metagalactic ionizing background at these redshifts. However, we caution that GN-UVC-1-like quasars are likely quite rare. The presence of such a LyC-luminous source in the relatively small comoving volume studied here is probably serendipitous, and a wider survey area (such as that used in CBT09) is needed to mitigate the effects of cosmic variance.

Meanwhile, for galaxies and low-luminosity AGNs like GN-UVC-6 to contribute significantly to the UV background, numerous fainter contributions would be required. We note, for example, that even with the high rate of contamination by foreground galaxies, Figure \ref{zspec_fig} only starts to become significantly populated at $z \gtrsim 2.4$ for apparent magnitudes approaching our cutoff of F275W = 26.

We may also think about the limits of our sample selection in terms of the UV continuum absolute magnitudes $M_{UV}$, usually measured at 1500 or 1600 \AA~when deriving rest-frame UV luminosity functions (LFs) at various redshifts. Again using the observed F606W magnitudes of our candidate sources as an estimate of the rest-frame 1500 \AA~flux, we find that our $z\sim3$ candidate LyC emitters probe as faint as $M_{UV} \approx$ -22.3. This is $\sim1.5$ magnitudes brighter than the characteristic luminosities of the $z\sim7$ rest-frame UV LFs derived in, e.g., \citet{bou15b} ($M^{*}_{1600}$ = -20.87) or \citet{liverm17} ($M^{*}_{1500}$ = -20.80). If our candidate LyC sources are taken to be analogs to the high-redshift galaxies that are responsible for reionization, these sources would thus still lie on the bright end of the $z\sim7$ UV LF. Further, even though the very deep \textit{HST} imaging used to construct these high-redshift UV LFs have detection limits as faint as $M_{UV} \sim -14.5$, even this is unable to detect the ultra-faint galaxies that appear to be required to complete hydrogen reionization by $z\sim6$ (e.g., \citealt{fink15,liverm17}). A more accurate census of $z\sim3$ analogs to the very-high-redshift sources that drove reionization will require deeper F275W imaging and corresponding spectroscopic follow-up, though at such faint magnitudes, redshift identifications are difficult.

\section{Summary}
\label{summary}
We have presented a search for candidate LyC emitters at $z \sim 3$ in the GOODS-North field using deep {\em HST}/WFC3 F275W imaging data and highly complete Keck/DEIMOS spectroscopic follow-up. We found five candidate ionizing sources brighter than F275W $= 26$, plus one additional source with blue F275W$-$F435W colors selected from a $B<25$ sample with colors $V-z_{850}<1$. One candidate (GN-UVC-1) is a $z \sim 2.5$ quasar which, at F275W $\sim 23.1$, is exceptionally bright at rest-frame wavelengths blueward of the Lyman limit. UV grism spectroscopy from \textit{HST}/WFC3 confirms the presence of significant LyC flux. Four candidates each appear to be contaminated by a foreground $z\sim0.5-0.7$ galaxy based on deep optical spectroscopy.

The contribution of the quasar GN-UVC-1 to the ionizing background at $z \sim 3$ totally dominates over the contributions from candidate LyC-emitting galaxies and faint AGNs (that is, GN-UVC-6, the sole non-quasar candidate source with no obvious contamination). Modulo potential currently-undetected contamination by lower redshift sources and the effects of cosmic variance, together they could account for the total ionizing background at $z\sim3$. However, for galaxies and low-luminosity AGNs alone to account for all (or even a non-negligible portion of) the total ionizing background at $z\sim3$, significant additional contributions from fainter sources would be needed. This will require deeper and wider area surveys to probe.

\acknowledgements
We gratefully acknowledge support from NASA grant NNX14AJ66G and NSF grant AST-1715145, the Trustees of the William F. Vilas Estate, and the University of Wisconsin-Madison Office of the Vice Chancellor for Research and Graduate Education with funding from the Wisconsin Alumni Research Foundation (A. J. B.). Support for program number GO-12479 was provided by NASA through a grant from the Space Telescope Science Institute, which is operated by the Association of Universities for Research in Astronomy, Inc., under NASA. Based in part on data obtained at the W. M. Keck
Observatory, which is operated as a scientific partnership among
the California Institute of Technology, the University of
California, and NASA and was made possible by the generous financial
support of the W. M. Keck Foundation.
The authors wish to recognize and acknowledge the very significant cultural role and reverence that the summit of Mauna Kea has always had within the indigenous Hawaiian community. We are most fortunate to have the opportunity to conduct observations from this mountain.

\end{document}